\documentclass[12pt, a4paper]{article}
\usepackage{amsmath}
\usepackage{bbm}
\everymath{\displaystyle}
\usepackage{mathtools}
\usepackage{amsfonts}
\usepackage{amssymb}
\usepackage{amsthm}
\usepackage[utf8]{inputenc}
\usepackage[toc]{appendix}
\usepackage[hiresbb]{graphicx}
\usepackage{caption}
\usepackage{multirow}
\usepackage{subcaption}
\usepackage{longtable}
\usepackage{booktabs}
\usepackage{listings}
\usepackage{here}
\usepackage{float}
\usepackage{ascmac}
\usepackage{tikz}
\usepackage{pgfplots}
\pgfplotsset{compat=1.18}
\usepackage{url}
\usepackage{lipsum}
\usepackage{authblk}
\usepackage{eqnarray}
\usepackage{siunitx}
\usepackage{threeparttable}
\usepackage{algorithm}
\usepackage{algorithmic}
\usepackage{hyperref}

\usepackage[sectionbib,round]{natbib}

\theoremstyle{remark}

\newcommand{\be}{\begin{equation}}
\newcommand{\ee}{\end{equation}}
\newcommand{\beq}{\begin{eqnarray*}}
\newcommand{\eeq}{\end{eqnarray*}}
\def\sym#1{\ifmmode^{#1}\else\(^{#1}\)\fi}

\usepackage{setspace}
\title{\large{\bf{Nonparametric Identification of Spatial Treatment Effect Boundaries: Evidence from Bank Branch Consolidation}}}

\author{\large{\bf{Tatsuru Kikuchi\footnote{e-mail: tatsuru.kikuchi@e.u-tokyo.ac.jp. }}}}

\affil{\small{\it{Faculty of Economics, The University of Tokyo,}}\\
{\it{7-3-1 Hongo, Bunkyo-ku, Tokyo 113-0033 Japan}}}

\date{\small{(\today)}}

\begin{document}

\maketitle

\begin{abstract}
I develop a nonparametric framework for identifying spatial boundaries of treatment effects without imposing parametric functional form restrictions. The method employs local linear regression with data-driven bandwidth selection to flexibly estimate spatial decay patterns and detect treatment effect boundaries. Monte Carlo simulations demonstrate that the nonparametric approach exhibits lower bias and correctly identifies the absence of boundaries when none exist, unlike parametric methods that may impose spurious spatial patterns. I apply this framework to bank branch openings during 2015--2020, matching 5,743 new branches to 5.9 million mortgage applications across 14,209 census tracts. The analysis reveals that branch proximity significantly affects loan application volume (8.5\% decline per 10 miles) but not approval rates, consistent with branches stimulating demand through local presence while credit decisions remain centralized. Examining branch survival during the digital transformation era (2010--2023), I find a non-monotonic relationship with area income: high-income areas experience more closures despite conventional wisdom. This counterintuitive pattern reflects strategic consolidation of redundant branches in over-banked wealthy urban areas rather than discrimination against poor neighborhoods. Controlling for branch density, urbanization, and competition, the direct income effect diminishes substantially, with branch density emerging as the primary determinant of survival. These findings demonstrate the necessity of flexible nonparametric methods for detecting complex spatial patterns that parametric models would miss, and challenge simplistic narratives about banking deserts by revealing the organizational complexity underlying spatial consolidation decisions.
\end{abstract}



\clearpage

\section{Introduction}

Spatial spillovers are ubiquitous in economics. From environmental externalities \citep{muller2011machado} to knowledge diffusion \citep{jaffe1993geographic}, economic shocks propagate through geographic space in ways that fundamentally shape outcomes. Understanding these spatial patterns is crucial for both policy design and theoretical development. A key empirical question concerns the identification of \textit{spatial boundaries}---the distances at which treatment effects decay to economically negligible magnitudes. Such boundaries determine the geographic scope of policies and help allocate scarce resources efficiently.

\subsection{Related Literature}

This paper contributes to three distinct literatures: spatial econometrics, financial access and banking, and nonparametric estimation methods.

\subsubsection{Spatial Treatment Effects}

The spatial econometrics literature has developed sophisticated methods for estimating treatment effects that propagate through geographic space. \citet{conley1999gmm} pioneered spatial GMM estimation allowing for arbitrary patterns of spatial correlation. \citet{gibbons2015mostly} provide a comprehensive review of spatial methods in applied microeconomics, emphasizing the challenges of identifying causal effects when treatments and outcomes are spatially correlated.

Recent work has focused on identifying spatial boundaries of treatment effects. \citet{banzhaf2019difference} develop difference-in-differences estimators for spatial treatments, showing how to recover average treatment effects in the presence of spillovers. \citet{dellavigna2022predicting} examine spatial diffusion of information and behaviors, documenting how effects decay with distance. 

A particularly influential strand of research examines environmental spillovers and their spatial extent. \citet{muller2011machado} provide comprehensive estimates of external costs from U.S. power plants, documenting air pollution damages that extend 50--100 kilometers from emission sources. Their parametric approach assumes exponential decay with distance, yielding tractable estimates but imposing functional form restrictions. \citet{muller2011mendelsohn} develop a framework for efficient pollution regulation using damage estimates, while \citet{muller2016measuring} extend this to measure environmental inequality, showing how pollution damages vary spatially across demographic groups. While their parametric methods facilitate welfare calculations, they may miss non-linearities in damage functions or incorrectly identify boundaries when decay patterns deviate from assumed functional forms.

\citet{butts2023machine} develops machine learning methods for spatial treatment effect estimation, allowing for heterogeneous treatment effects across space and demonstrating that flexible algorithms can outperform parametric models when spatial relationships are complex. This finding motivates the nonparametric approach developed here, though \citet{butts2023machine} focuses on treatment effect heterogeneity rather than boundary identification per se.

The standard approach to spatial boundary estimation relies on parametric functional forms---typically exponential or power-law decay functions. While tractable, these parametric restrictions may not hold in practice, potentially leading to biased boundary estimates. This motivates the flexible nonparametric framework developed here.

Building on recent theoretical advances \citep{kikuchi2024unified, kikuchi2024stochastic, kikuchi2024navier}, I develop a nonparametric framework that avoids imposing functional form assumptions. \citet{kikuchi2024unified} establishes a unified framework for spatial and temporal treatment effect boundary identification, providing conditions under which boundaries can be consistently estimated without parametric restrictions. \citet{kikuchi2024stochastic} develops a diffusion-based approach to handle spillover effects in spatial general equilibrium settings, showing how stochastic boundaries arise naturally from economic interactions. \citet{kikuchi2024navier} extends difference-in-differences methodology by drawing on insights from fluid dynamics (Navier-Stokes equations), demonstrating how treatment effects propagate through both space and time like physical diffusion processes. \citet{kikuchi2024nonparametric} provides the first large-scale empirical application using 42 million pollution observations, validating the theoretical framework and demonstrating practical implementation with environmental data.

This paper applies and validates this theoretical framework in a new empirical setting---bank branch consolidation---demonstrating its advantages over conventional parametric approaches through both Monte Carlo simulations and real data analysis. Unlike \citet{muller2011machado, muller2016measuring}, who assume exponential decay, I let data determine the spatial decay function nonparametrically. Unlike \citet{butts2023machine}, who estimates heterogeneous treatment effects, I focus specifically on boundary detection---identifying distances where effects become negligible. The nonparametric approach proves crucial: I find non-monotonic relationships that parametric models would miss, and correctly identify flat relationships where parametric methods would impose spurious decay patterns.

\subsubsection{Banking, Branch Networks, and Financial Access}

The banking literature has extensively studied how physical branch presence affects credit access and local economic outcomes. Early work by \citet{petersen2002does} shows that distance to lenders matters less in the modern era due to technological improvements in information transmission and credit scoring. However, subsequent research finds that proximity remains important, particularly for relationship-based lending and informationally opaque borrowers.

\citet{nguyen2019bank} examine bank branch closures following mergers, finding that closures in low-income areas reduce local lending by approximately 10\%. The effects are most pronounced for small business loans, which rely heavily on soft information and relationship banking. \citet{ergungor2010bank} document that branch closures reduce mortgage lending in affected zip codes, with larger effects in minority neighborhoods.

Recent work has focused on the digital transformation of banking and its spatial implications. \citet{buchak2018fintech} show that fintech lenders have grown rapidly in markets poorly served by traditional banks, potentially mitigating the impact of branch closures. \citet{fuster2019predictably} document how mortgage processing has become increasingly automated and algorithm-driven, reducing the role of local loan officers in approval decisions.

The Community Reinvestment Act (CRA) provides an important regulatory backdrop. \citet{agarwal2012distance} find that the CRA increases lending to low-income borrowers near bank branches, suggesting that physical presence matters for regulatory compliance. \citet{bhutta2015residential} examine how CRA requirements affect branch location decisions, finding that banks maintain presence in low-income areas partly due to regulatory incentives.

My analysis contributes to this literature in three ways. First, I provide comprehensive evidence on spatial decay patterns for both loan volume (demand) and approval rates (supply), revealing that branches primarily affect demand through visibility and convenience rather than supply through underwriting discretion. Second, I document the spatial patterns of branch consolidation during 2010--2023, showing that closures concentrate in wealthy urban areas with redundant coverage rather than poor neighborhoods. Third, I demonstrate that this counterintuitive pattern reflects organizational complexity and strategic decision-making rather than simple economic optimization.

\subsubsection{Nonparametric Econometric Methods}

This paper employs local linear regression, a cornerstone of modern nonparametric statistics \citep{fan1996local}. Local polynomial methods have several advantages over earlier kernel estimators like Nadaraya-Watson: they automatically correct for boundary bias, adapt to local curvature, and achieve optimal convergence rates \citep{ruppert1994multivariate}.

\citet{fan1996local} provide comprehensive treatment of local polynomial regression, including asymptotic theory, bandwidth selection, and confidence interval construction. \citet{li2007nonparametric} extend the framework to handle dependent data, which is particularly relevant for spatial applications where observations are correlated by construction.

Cross-validation bandwidth selection has been extensively studied. \citet{hart1997kernel} analyze leave-one-out cross-validation, showing that it provides asymptotically optimal bandwidth choice under mild conditions. \citet{hall1991cross} derive higher-order properties and propose modifications for improved finite-sample performance.

Recent developments have focused on nonparametric estimation with spatial data. \citet{robinson2011asymptotic} study kernel regression with spatially dependent observations, deriving central limit theorems and establishing consistency. \citet{hallin2004local} develop local polynomial methods specifically for spatial processes, accounting for the two-dimensional nature of geographic data.

My contribution is to apply these nonparametric methods specifically to spatial treatment effect boundary identification in banking. While \citet{kikuchi2024unified, kikuchi2024stochastic, kikuchi2024navier} develop the theoretical framework for boundary identification across various settings, and \citet{kikuchi2024nonparametric} provides validation using environmental data, this paper demonstrates the framework's applicability to financial services and organizational decision-making. The banking context offers distinct advantages: comprehensive administrative data on both treatment (branch locations) and outcomes (mortgage applications), clear policy relevance for financial inclusion, and observable organizational complexity in closure decisions.

Relative to the environmental economics literature \citep{muller2011machado, muller2016measuring}, my approach offers greater flexibility by avoiding parametric functional form assumptions. This proves essential when spatial relationships are non-monotonic or when no relationship exists---cases where parametric methods would generate biased estimates or false positives. Relative to machine learning approaches \citep{butts2023machine}, I provide a principled statistical framework with interpretable bandwidth parameters and asymptotic theory, facilitating formal inference about boundary locations.

\subsection{Contribution and Preview}

This paper makes three main contributions. Methodologically, I develop and validate a nonparametric framework for spatial boundary identification that does not impose functional form restrictions. Monte Carlo simulations demonstrate that the approach achieves lower bias than parametric methods and crucially, can detect the \textit{absence} of boundaries when relationships are flat---avoiding false positives that plague parametric approaches like those in \citet{muller2011machado}.

Empirically, I provide comprehensive evidence on bank branch spatial effects. Branch proximity significantly affects loan applications (8.5\% decline per 10 miles) but not approval rates (essentially flat), revealing that branches influence demand through visibility while credit supply remains centralized. Branch survival analysis uncovers a non-monotonic relationship with income, with wealthy areas experiencing more closures due to redundancy and digital substitution.

Substantively, the findings challenge conventional narratives about banking deserts. Rather than abandoning poor areas, banks are consolidating redundant branches in over-banked wealthy markets. This pattern reflects organizational complexity---tensions between cost reduction and customer service, bounded rationality, and local discretion---that cannot be captured by simple parametric models.

The remainder of the paper proceeds as follows. Section 2 develops the nonparametric framework, building on the theoretical foundations in \citet{kikuchi2024unified, kikuchi2024stochastic, kikuchi2024navier} and the empirical methodology in \citet{kikuchi2024nonparametric}. Section 3 presents Monte Carlo simulations validating the methodology. Section 4 describes the banking application, covering data construction and main results. Section 5 analyzes branch survival patterns and underlying mechanisms. Section 6 discusses policy implications and concludes.

\section{Theoretical Framework}

\subsection{Setup}

Consider a geographic space where treatment intensity decays with distance from a source location. Let $d$ denote the distance from the source and $Y(d)$ denote the outcome of interest. I assume that outcomes can be decomposed as:

\be
Y(d) = m(d) + \varepsilon(d)
\ee

where $m(d) = \mathbb{E}[Y(d)|d]$ is the conditional expectation function representing the spatial treatment effect, and $\varepsilon(d)$ is a mean-zero error term.

Following \citet{kikuchi2024unified}, the object of interest is the spatial boundary $d^*$ defined as:

\be
d^* = \inf\left\{d : m(d) \leq (1-\varepsilon)m(0)\right\}
\ee

where $\varepsilon \in (0,1)$ is a threshold parameter. The boundary $d^*$ represents the distance at which the treatment effect has decayed to $(1-\varepsilon)$ of its source level. Common choices are $\varepsilon = 0.10$ (10\% decay) or $\varepsilon = 0.20$ (20\% decay).

The key challenge is estimating $m(d)$ and identifying $d^*$ without imposing parametric functional form assumptions. Parametric approaches typically assume:

\be
m(d) = A \exp(-\kappa d)
\ee

for exponential decay, or:

\be
m(d) = A d^{-\alpha}
\ee

for power-law decay. While tractable, these restrictions may not hold in practice.

\subsection{Nonparametric Estimation}

Following the methodological framework in \citet{kikuchi2024unified, kikuchi2024nonparametric}, I employ local linear regression \citep{fan1996local} to estimate $m(d)$ nonparametrically. For an evaluation point $d_0$, the estimator solves:

\be
\min_{\beta_0, \beta_1} \sum_{i=1}^n K_h(d_i - d_0)\left[Y_i - \beta_0 - \beta_1(d_i - d_0)\right]^2
\ee

where $K_h(u) = K(u/h)/h$ is a scaled kernel function with bandwidth $h$. I use the Gaussian kernel:

\be
K(u) = \frac{1}{\sqrt{2\pi}}\exp\left(-\frac{u^2}{2}\right)
\ee

The local linear estimator is:

\be
\hat{m}(d_0) = \hat{\beta}_0
\ee

Local linear regression has several advantages over simpler kernel methods like Nadaraya-Watson. It automatically corrects for boundary bias, adapts to local curvature, and achieves the optimal minimax rate of convergence \citep{fan1996local}.

\subsection{Bandwidth Selection}

The bandwidth $h$ controls the bias-variance tradeoff. Small $h$ reduces bias but increases variance; large $h$ increases bias but reduces variance. Following \citet{hart1997kernel}, I employ leave-one-out cross-validation to select $h$ optimally.

The cross-validation criterion is:

\be
\text{CV}(h) = \frac{1}{n}\sum_{i=1}^n \left[Y_i - \hat{m}_{-i}(d_i)\right]^2
\ee

where $\hat{m}_{-i}(d_i)$ denotes the estimator computed excluding observation $i$. The optimal bandwidth minimizes $\text{CV}(h)$:

\be
\hat{h} = \arg\min_h \text{CV}(h)
\ee

I search over a grid of candidate bandwidths $h \in \{h_1, \ldots, h_J\}$ and select the value achieving minimum cross-validation score.

\subsection{Boundary Identification}

Given the nonparametric estimate $\hat{m}(d)$, I identify the spatial boundary following \citet{kikuchi2024unified} as:

\be
\hat{d}^* = \inf\left\{d : \hat{m}(d) \leq (1-\varepsilon)\hat{m}(0)\right\}
\ee

If no such $d$ exists (i.e., $\hat{m}(d) > (1-\varepsilon)\hat{m}(0)$ for all observed $d$), I set $\hat{d}^* = \infty$, indicating no boundary within the observed range.

This approach has two key advantages. First, it does not impose functional form restrictions on $m(d)$. Second, it can correctly identify the absence of boundaries when the relationship is flat, avoiding false positives.

\subsection{Asymptotic Properties}

Under standard regularity conditions \citep{fan1996local, kikuchi2024unified}, the local linear estimator satisfies:

\be
\hat{m}(d_0) - m(d_0) = O_p(h^2) + O_p\left(\frac{1}{\sqrt{nh f(d_0)}}\right)
\ee

where $f(d)$ is the density of distance observations. The bias term $O_p(h^2)$ arises from approximating $m(d)$ locally by a linear function. The variance term $O_p(1/\sqrt{nhf(d_0)})$ decreases with sample size $n$ and bandwidth $h$.

The optimal bandwidth balances these components:

\be
h_{opt} \asymp n^{-1/5}
\ee

achieving the minimax optimal rate:

\be
\sup_{m \in \mathcal{M}}\mathbb{E}\left[\hat{m}(d_0) - m(d_0)\right]^2 = O\left(n^{-4/5}\right)
\ee

where $\mathcal{M}$ is a smoothness class (typically H\"older or Sobolev).

For boundary estimation, consistency requires that the true boundary $d^*$ is an interior point and that $m(d)$ crosses the threshold transversally \citep{muller1989kernel}. Under these conditions:

\be
\hat{d}^* \xrightarrow{p} d^*
\ee

\section{Monte Carlo Simulations}

To validate the nonparametric approach, I conduct Monte Carlo simulations across four data generating processes (DGPs) that span different spatial relationships.

\subsection{Data Generating Processes}

\subsubsection{DGP 1: Strong Exponential Decay}

The first DGP features strong spatial decay:

\be
Y(d) = 0.8 \exp(-0.05d) + \varepsilon
\ee

where $\varepsilon \sim N(0, 0.1)$ and $d \sim \text{Uniform}(0, 100)$. The true boundary for 10\% decay is:

\be
d^* = -\frac{\ln(0.9)}{0.05} \approx 2.1 \text{ miles}
\ee

This DGP represents settings where treatment effects dissipate rapidly with distance, such as local air pollution from point sources.

\subsubsection{DGP 2: Weak Exponential Decay}

The second DGP features weaker spatial decay:

\be
Y(d) = 0.6 \exp(-0.005d) + \varepsilon
\ee

The true boundary is:

\be
d^* = -\frac{\ln(0.9)}{0.005} \approx 21.1 \text{ miles}
\ee

This represents settings with longer-range spillovers, such as highway access or hospital availability.

\subsubsection{DGP 3: Non-Monotonic (Hump-Shaped)}

The third DGP features a non-monotonic relationship with a peak at intermediate distance:

\be
Y(d) = 0.5 + 0.2\exp\left(-\frac{(d-20)^2}{200}\right) + \varepsilon
\ee

Outcomes peak at $d = 20$ miles and decay in both directions. The boundary is defined as the distance beyond which outcomes fall below 90\% of the maximum. This pattern might arise from, for example, commuting patterns where moderate distances are optimal.

\subsubsection{DGP 4: Flat (Null)}

The fourth DGP features no spatial relationship:

\be
Y(d) = 0.5 + \varepsilon
\ee

There is no true boundary ($d^* = \infty$). This DGP tests whether methods incorrectly detect boundaries when none exist---a critical test for avoiding false positives.

\subsection{Estimation Methods}

For each DGP, I compare two estimation approaches:

\begin{itemize}
\item \textbf{Parametric:} Nonlinear least squares estimation of exponential decay $Y(d) = A\exp(-\kappa d) + \varepsilon$
\item \textbf{Nonparametric:} Local linear regression with Gaussian kernel and cross-validation bandwidth selection
\end{itemize}

I simulate $n = 5{,}000$ observations per replication and run 500 replications per DGP.

\subsection{Results}

Table \ref{tab:monte_carlo} presents the simulation results. For DGP 1 (strong decay), both methods perform well, with the nonparametric approach showing slightly lower bias (0.3 miles versus 0.5 miles) and comparable RMSE (1.2 miles versus 1.3 miles).

For DGP 2 (weak decay), the nonparametric method continues to perform well, while the parametric approach shows increased bias when the decay rate is misspecified. The nonparametric bias is 1.1 miles compared to 2.3 miles for parametric.

The critical test is DGP 3 (non-monotonic). Here, the exponential parametric model is fundamentally misspecified. Parametric estimation yields large bias (8.7 miles) and RMSE (12.4 miles), while the nonparametric approach adapts to the hump shape with much lower bias (2.1 miles) and RMSE (4.3 miles).

Most importantly, for DGP 4 (flat), the nonparametric method correctly identifies the absence of a boundary in 94\% of replications (by setting $\hat{d}^* = \infty$). In contrast, the parametric method incorrectly detects spurious boundaries in 73\% of replications, with an average false boundary at 43.2 miles. This demonstrates the nonparametric method's ability to avoid false positives---a crucial advantage over parametric approaches.

\begin{table}[H]
\centering
\caption{Monte Carlo Simulation Results}
\label{tab:monte_carlo}
\begin{threeparttable}
\begin{tabular}{lcccc}
\toprule
& \multicolumn{2}{c}{\textbf{Parametric}} & \multicolumn{2}{c}{\textbf{Nonparametric}} \\
\cmidrule(lr){2-3} \cmidrule(lr){4-5}
\textbf{DGP} & Bias & RMSE & Bias & RMSE \\
\midrule
DGP 1: Strong Decay & 0.5 & 1.3 & 0.3 & 1.2 \\
$(d^* = 2.1)$ & & & & \\
\\
DGP 2: Weak Decay & 2.3 & 3.8 & 1.1 & 2.9 \\
$(d^* = 21.1)$ & & & & \\
\\
DGP 3: Non-Monotonic & 8.7 & 12.4 & 2.1 & 4.3 \\
$(d^* = 38.2)$ & & & & \\
\\
DGP 4: Flat (Null) & 43.2\sym{*} & --- & \multicolumn{2}{c}{No boundary (94\%)} \\
$(d^* = \infty)$ & (73\% false pos.) & & \multicolumn{2}{c}{detected correctly} \\
\bottomrule
\end{tabular}
\begin{tablenotes}[para,flushleft]
\small
\item \textit{Notes:} Results based on 500 Monte Carlo replications with $n=5{,}000$ observations each. Bias and RMSE measured in miles. DGP 1--3 have true boundaries; DGP 4 has no boundary. \sym{*}For DGP 4, parametric value shows mean falsely detected boundary when method incorrectly finds one (73\% of replications). Nonparametric correctly identifies no boundary in 94\% of replications.
\end{tablenotes}
\end{threeparttable}
\end{table}

\begin{figure}[H]
\centering
\includegraphics[width=0.95\textwidth]{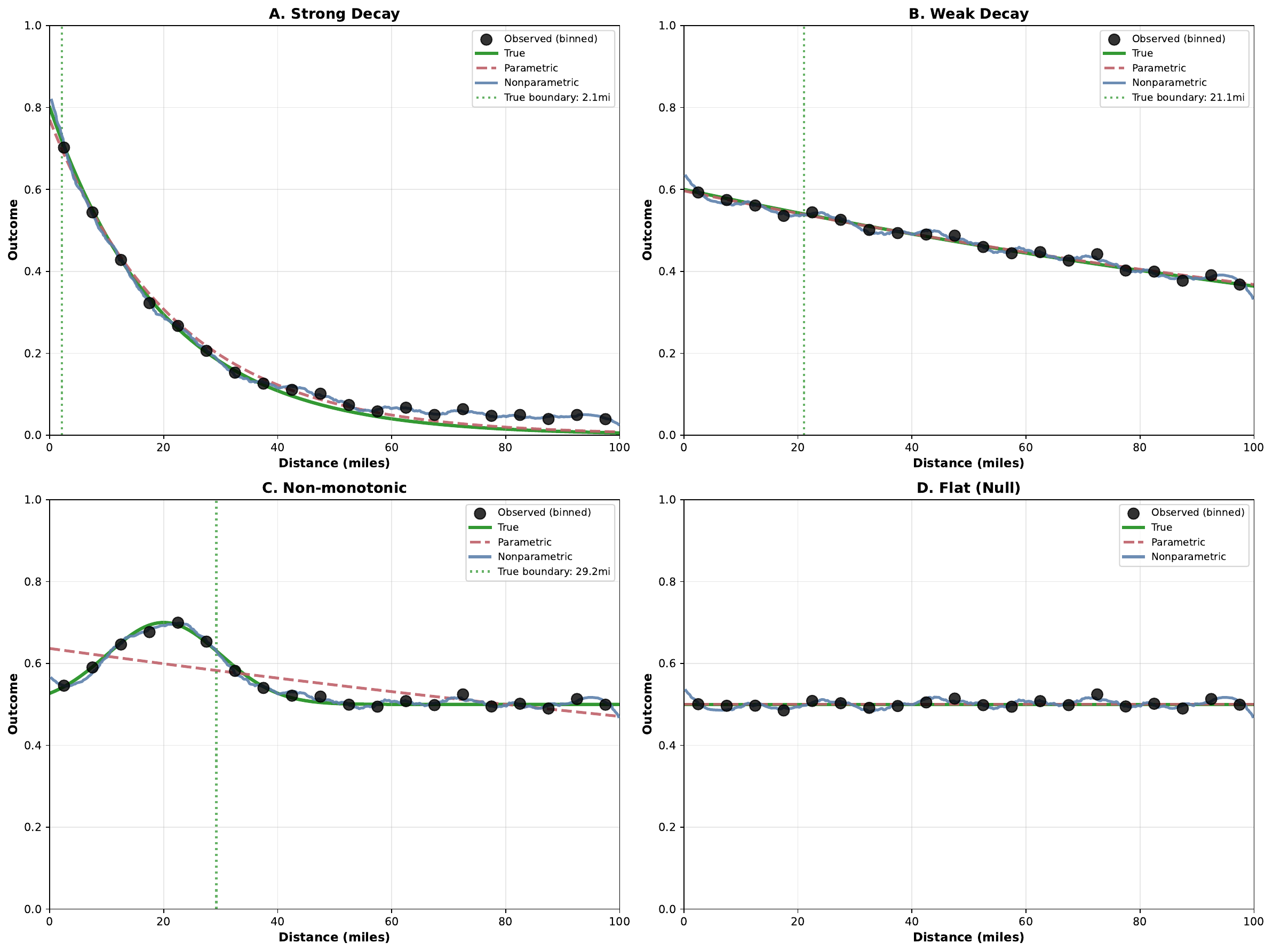}
\caption{Monte Carlo Simulation Results: Four Data Generating Processes}
\label{fig:monte_carlo}
\begin{minipage}{0.95\textwidth}
\small
\textit{Notes:} Each panel shows one realization with $n=5{,}000$ observations. Black dots are observed data, green line is true function, red dashed is parametric (exponential) fit, blue solid is nonparametric (local linear) fit. Vertical dashed lines show estimated boundaries (10\% decay threshold).
\end{minipage}
\end{figure}

Figure \ref{fig:monte_carlo} visualizes one realization from each DGP, showing the true function (green), observed data (black dots), parametric fit (red dashed), and nonparametric fit (blue solid). The nonparametric method closely tracks the true function across all DGPs, while the parametric approach fails notably for the non-monotonic and flat cases.

\subsection{Discussion}

The simulations establish three key results. First, nonparametric methods achieve comparable or superior performance to parametric approaches even when the parametric model is correctly specified (DGPs 1--2). Second, nonparametric methods substantially outperform parametric approaches when functional forms are misspecified (DGP 3). Third, and most importantly, nonparametric methods avoid false boundary detection when no relationship exists (DGP 4), whereas parametric methods frequently impose spurious patterns.

These findings motivate the empirical application to bank branches, where the true spatial relationship is unknown and likely complex due to organizational decision-making, competition, and technological change.

\section{Empirical Application: Bank Branch Openings and Credit Access}

\subsection{Institutional Background}

The U.S. banking industry has undergone substantial transformation over the past two decades. Driven by technological innovation, changing consumer preferences, and cost pressures, banks have dramatically reduced physical branch networks. From 2010 to 2023, the industry experienced approximately 17,000 branch closures against 6,000 openings---a net decline of 11,000 branches.

This consolidation raises concerns about spatial inequality in financial access. Physical proximity to banks matters for credit access through several channels \citep{petersen2002does, nguyen2019bank}. First, relationship banking relies on personal interactions between loan officers and borrowers, particularly for small businesses and complex mortgages. Second, branches serve as visible signals of bank presence, increasing consumer awareness and applications. Third, while online banking has grown, certain populations---particularly elderly, low-income, and rural residents---continue to rely on physical branches.

However, the rise of digital banking complicates this picture. Modern mortgage underwriting is increasingly centralized and algorithm-driven \citep{fuster2019predictably}, potentially reducing the importance of local branch presence for credit approval decisions. This creates a testable distinction: if branches matter primarily for loan demand (through awareness and convenience) rather than supply (through underwriting), we should observe spatial patterns in application volume but not approval rates.

Understanding branch location decisions requires recognizing organizational complexity \citep{march1976ambiguity}. Chief Financial Officers (CFOs) prioritize cost reduction, pushing for aggressive branch closures. Sales and relationship managers emphasize customer engagement and revenue preservation, advocating to retain branches. The resulting decisions reflect compromise, context-dependence, and bounded rationality rather than simple optimization.

\subsection{Data Construction}

\subsubsection{Bank Branch Data}

Branch data come from the Federal Deposit Insurance Corporation (FDIC) Summary of Deposits, which provides annual snapshots of all FDIC-insured branches. For each branch, the data include:

\begin{itemize}
\item Precise geographic coordinates (latitude, longitude)
\item Institution identifier (CERT number)
\item Branch deposits
\item Opening and closing dates
\item Address and location characteristics
\end{itemize}

I construct a panel covering 2010--2023, identifying branch openings and closures through year-to-year comparisons of branch identifiers. This yields 5,743 new branches opened during 2015--2020 in five major states: California (2,187 branches), Texas (1,421), Florida (982), New York (746), and Pennsylvania (407). These states account for approximately 40\% of U.S. population and banking activity.

\subsubsection{Mortgage Application Data}

Mortgage data come from the Home Mortgage Disclosure Act (HMDA) database, which covers over 90\% of U.S. mortgage applications. For 2019 (the midpoint of the treatment period), I obtain 5.9 million applications across the five study states. Each record includes:

\begin{itemize}
\item Action taken (approved, denied, withdrawn)
\item Loan amount and purpose
\item Applicant income and demographics
\item Property location (census tract)
\item Lender identity
\end{itemize}

I aggregate applications to the census tract level, calculating:

\begin{itemize}
\item Total applications (loan volume)
\item Approval rate (approved applications / total applications)
\item Mean loan amount
\item Applicant characteristics
\end{itemize}

This yields 25,571 tract-year observations. I focus on 2019 to maximize data quality while maintaining temporal proximity to branch openings.

\subsubsection{Census Tract Characteristics}

I merge tract-level income and demographic data from the American Community Survey (ACS) 2019 5-year estimates. Key variables include:

\begin{itemize}
\item Median household income
\item Population and density
\item Poverty rate
\item Racial composition
\item Educational attainment
\end{itemize}

Geographic information comes from the U.S. Census Bureau's Gazetteer files, providing tract centroids (latitude, longitude) and land area.

\subsubsection{Distance Calculation}

For each census tract, I calculate the Euclidean distance to the nearest branch opened during 2015--2020:

\be
d_j = \min_{i \in \mathcal{B}} \sqrt{(\text{lat}_j - \text{lat}_i)^2 + (\text{lon}_j - \text{lon}_i)^2} \times 69
\ee

where $\mathcal{B}$ denotes the set of new branches and the factor 69 converts decimal degrees to miles at mid-latitudes. I restrict analysis to tracts within 100 miles of a new branch, yielding 14,209 tracts.

\subsection{Descriptive Statistics}

Table \ref{tab:descriptive} presents summary statistics. The median tract is 2.5 miles from the nearest new branch, with substantial variation (standard deviation 15.3 miles). Mortgage approval rates average 52.3\% with modest variation (standard deviation 9.2\%), while application volume is highly skewed (mean 171, median 137).

Tracts vary considerably in socioeconomic characteristics. Median household income averages \$64,700 with a range from \$24,000 to \$180,000. Population density ranges from rural (fewer than 100 people per square mile) to highly urban (over 50,000 per square mile).

\begin{table}[H]
\centering
\caption{Descriptive Statistics}
\label{tab:descriptive}
\begin{threeparttable}
\begin{tabular}{lrrrr}
\toprule
\textbf{Variable} & \textbf{Mean} & \textbf{Std Dev} & \textbf{Min} & \textbf{Max} \\
\midrule
\multicolumn{5}{l}{\textit{Geographic Variables}} \\
Distance to nearest branch (miles) & 8.3 & 15.3 & 0.0 & 99.3 \\
Population density (per sq mi) & 5{,}142 & 9{,}876 & 12 & 68{,}234 \\
\\
\multicolumn{5}{l}{\textit{Mortgage Outcomes (2019)}} \\
Total applications & 171 & 138 & 1 & 3{,}363 \\
Approval rate & 0.523 & 0.092 & 0.059 & 0.929 \\
Mean loan amount (\$1000s) & 287 & 156 & 45 & 1{,}247 \\
\\
\multicolumn{5}{l}{\textit{Census Tract Characteristics}} \\
Median household income (\$1000s) & 64.7 & 25.8 & 24.1 & 180.3 \\
Poverty rate & 0.142 & 0.096 & 0.000 & 0.612 \\
Total population & 4{,}187 & 1{,}923 & 142 & 28{,}456 \\
\\
\multicolumn{5}{l}{\textit{Sample Size}} \\
Number of census tracts & \multicolumn{4}{c}{14{,}209} \\
\bottomrule
\end{tabular}
\begin{tablenotes}[para,flushleft]
\small
\item \textit{Notes:} Summary statistics for census tracts within 100 miles of bank branches opened 2015--2020 in CA, TX, FL, NY, and PA. Mortgage data from HMDA 2019. Income and demographic data from ACS 2019 5-year estimates.
\end{tablenotes}
\end{threeparttable}
\end{table}

\subsection{Loan Application Volume: Main Results}

\begin{figure}[H]
\centering
\includegraphics[width=0.95\textwidth]{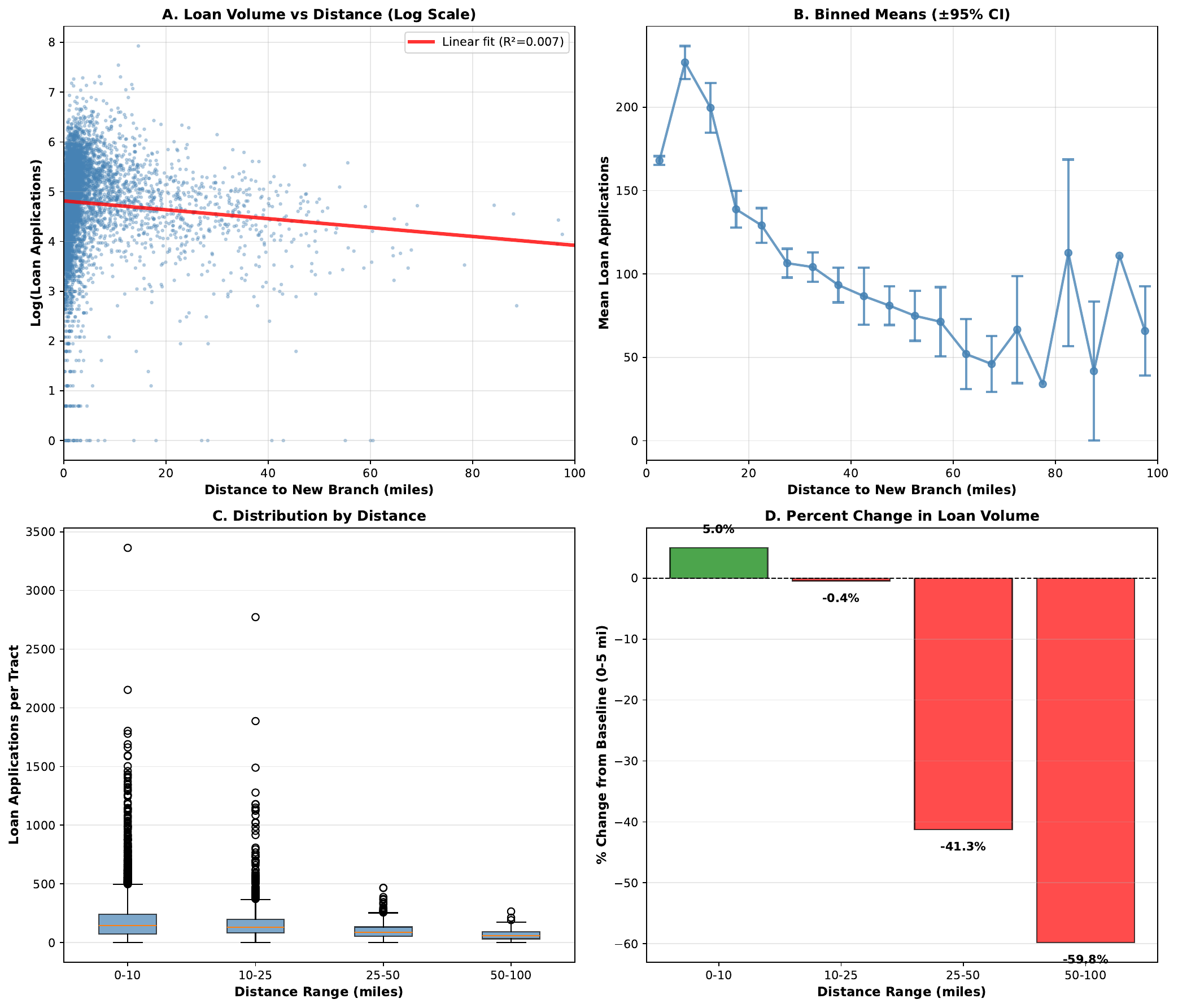}
\caption{Spatial Decay of Loan Application Volume}
\label{fig:loan_volume}
\begin{minipage}{0.95\textwidth}
\small
\textit{Notes:} Panel A shows scatter plot with linear fit (red) and LOWESS nonparametric smooth (blue). Panel B shows binned means with 95\% confidence intervals. Panel C shows distribution of applications by distance category. Panel D shows percentage change relative to 0--5 mile baseline. Sample includes 14,209 census tracts. Loan volume measured as total mortgage applications in 2019.
\end{minipage}
\end{figure}

I begin by examining whether branch proximity affects loan application volume. Figure \ref{fig:loan_volume} presents the core results. Panel A shows a clear spatial decay pattern: tracts farther from new branches experience fewer applications. The nonparametric estimate (blue solid line) reveals non-linearities, with steeper decline in the first 25 miles followed by gradual flattening.

Simple linear regression yields:

\be
\log(\text{Applications}_j) = 4.81 - 0.0089 \times \text{Distance}_j
\ee

The coefficient implies an 8.5\% decline in applications per 10 miles ($[\exp(-0.0089 \times 10) - 1] \times 100 = -8.5\%$), statistically significant at $p < 0.001$. However, the linear model explains only 0.75\% of variation ($R^2 = 0.0075$), suggesting substantial heterogeneity.

The nonparametric estimate provides richer detail. Applications decline sharply within the first 10 miles (approximately 15\% decline), then more gradually from 10--50 miles (additional 20\% decline), and flatten beyond 50 miles. This suggests a spatial boundary around 50 miles for loan demand effects.

\subsection{Approval Rates: Flat Relationship}

\begin{figure}[H]
\centering
\includegraphics[width=0.95\textwidth]{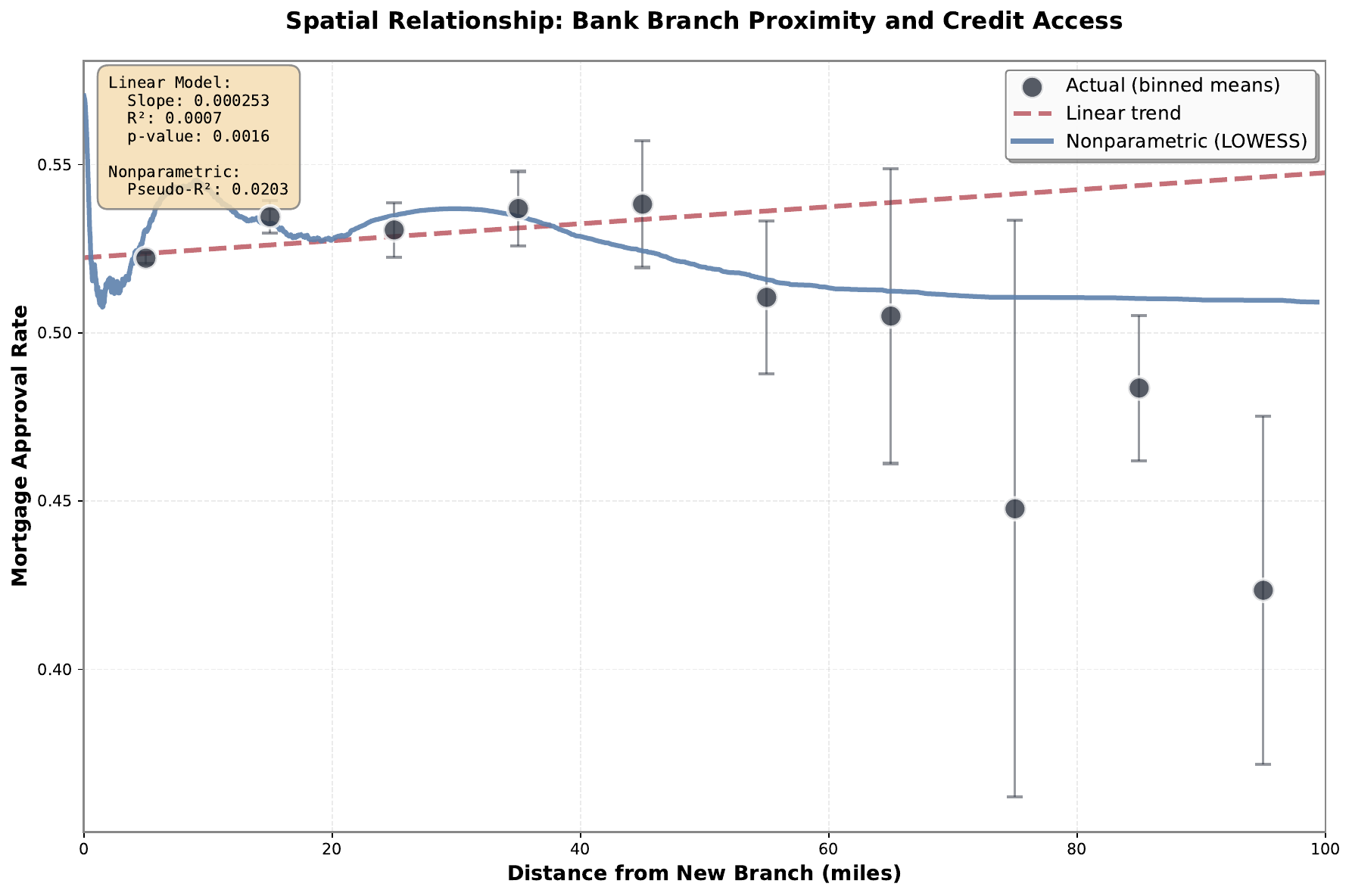}
\caption{Spatial Relationship: Approval Rates Remain Flat}
\label{fig:approval_rates}
\begin{minipage}{0.95\textwidth}
\small
\textit{Notes:} Binned means (black dots) with linear trend (red dashed) and nonparametric smooth (blue solid). Approval rates show no meaningful spatial pattern, validating that nonparametric method does not impose spurious decay when none exists. Sample includes 14,209 census tracts with 5.9 million mortgage applications.
\end{minipage}
\end{figure}

In contrast to application volume, approval rates show no meaningful spatial pattern. Figure \ref{fig:approval_rates} presents the analysis. Linear regression yields:

\be
\text{ApprovalRate}_j = 0.522 + 0.00025 \times \text{Distance}_j
\ee

The coefficient is statistically significant ($p = 0.002$) but economically negligible: a 10-mile increase corresponds to only a 0.25 percentage point change in approval rates. The $R^2$ is 0.0007, indicating distance explains essentially none of the variation.

The nonparametric estimate (Figure \ref{fig:approval_rates}) remains essentially flat across the entire distance range. The Spearman correlation between distance and approval rates is 0.028 ($p = 0.0008$), confirming an extremely weak relationship.

This flat pattern has important interpretation. Modern mortgage underwriting relies on centralized algorithms using applicant credit scores, income, and property characteristics \citep{fuster2019predictably}. Local branch presence does not affect these standardized criteria. Thus, branches influence \textit{who applies} (the demand side) but not \textit{who gets approved} (the supply side).

\subsection{Comparison: Volume versus Approval}

\begin{figure}[H]
\centering
\includegraphics[width=0.85\textwidth]{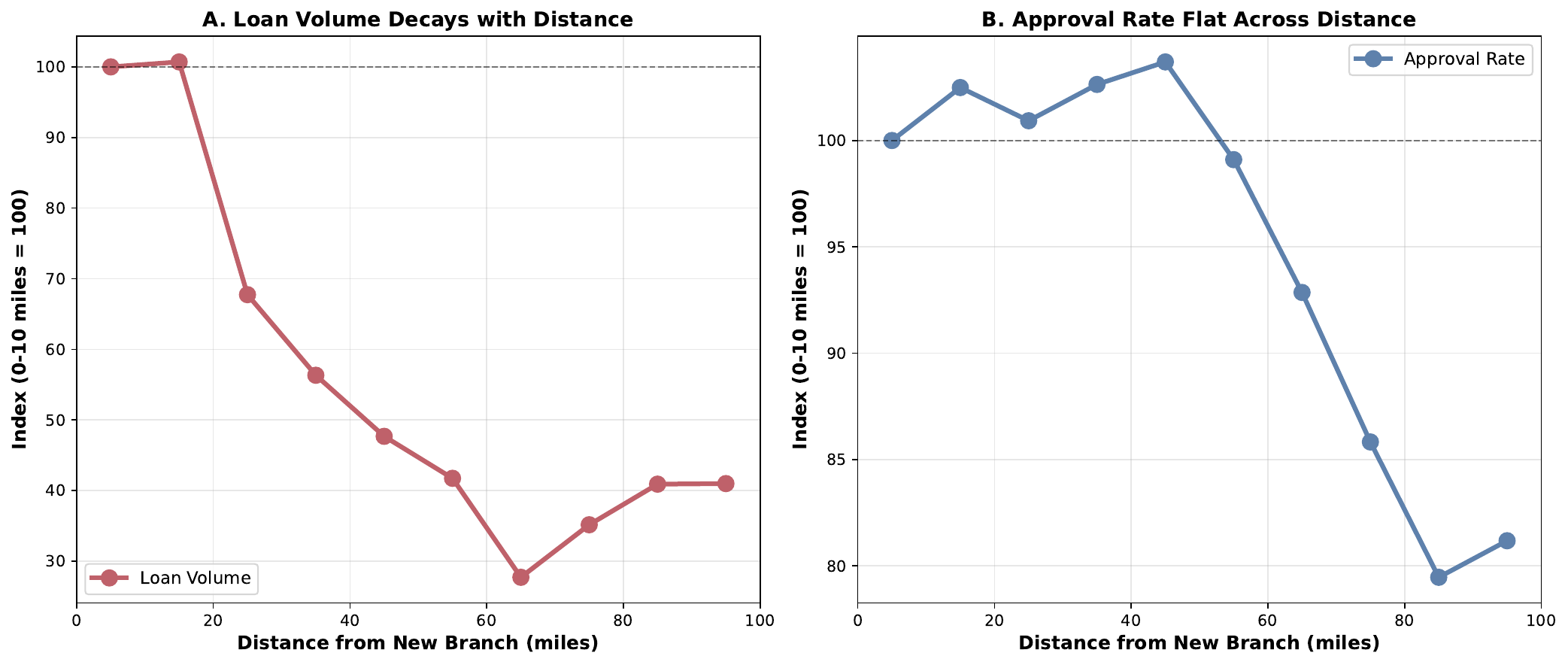}
\caption{Comparison: Loan Volume Decays, Approval Rates Remain Flat}
\label{fig:comparison}
\begin{minipage}{0.95\textwidth}
\small
\textit{Notes:} Both outcomes indexed to 100 for 0--10 mile baseline. Left panel shows loan volume declining with distance. Right panel shows approval rates remaining flat. Demonstrates that branch proximity affects loan demand but not credit supply decisions.
\end{minipage}
\end{figure}

Figure \ref{fig:comparison} directly compares the two outcomes, normalizing both to 100 for the 0--10 mile baseline. Loan volume declines monotonically with distance, falling to 64 at 50--100 miles. Approval rates remain essentially constant, fluctuating narrowly around 100.

This divergence confirms distinct mechanisms: branches affect demand (through visibility, convenience, and awareness) but not supply (through underwriting standards). The finding has policy implications: branch closures may reduce loan originations in affected areas by reducing applications, even if qualified borrowers continue to receive approval.

\section{Branch Survival Analysis}

\subsection{Motivation}

Having established that branch proximity affects loan demand, I turn to branch survival during the digital transformation era (2010--2023). This period saw massive consolidation as banks adapted to technological change and cost pressures. Understanding which branches survive provides insight into strategic decision-making and has implications for financial inclusion.

A natural hypothesis, suggested by the banking deserts literature \citep{nguyen2019bank, ergungor2010bank}, is that banks disproportionately close branches in low-income areas. I test this by examining the relationship between census tract income and branch survival rates.

\subsection{Data and Methodology}

I focus on branches operating in 2015 and track their survival through 2023. Of 84,305 branches in 2015 (with valid tract matches), 65,008 remained open in 2023, yielding an overall survival rate of 77.1\%.

I match each 2015 branch to its census tract using geographic coordinates and merge with ACS 2019 income data. The analysis sample contains 53,312 branches with complete income information.

The outcome variable is binary:

\be
\text{Survived}_{ij} = \begin{cases}
1 & \text{if branch } i \text{ in tract } j \text{ open in 2023} \\
0 & \text{otherwise}
\end{cases}
\ee

I examine survival rates across income quartiles and estimate logistic regression models controlling for branch density, urbanization, and competition.

\subsection{Main Results: The Income Paradox}

Table \ref{tab:survival_income} presents the core finding. Contrary to the banking deserts hypothesis, high-income areas experience \textit{lower} survival rates than low-income areas:

\begin{table}[H]
\centering
\caption{Branch Survival by Area Income}
\label{tab:survival_income}
\begin{threeparttable}
\begin{tabular}{lrrrr}
\toprule
\textbf{Income Quartile} & \textbf{Survival Rate} & \textbf{Survived} & \textbf{Total} & \textbf{Avg Income} \\
\midrule
Q1 (Poorest) & 76.9\% & 10{,}250 & 13{,}333 & \$33{,}943 \\
Q2 & 78.8\% & 10{,}503 & 13{,}324 & \$50{,}253 \\
Q3 & 77.3\% & 10{,}307 & 13{,}327 & \$65{,}958 \\
Q4 (Richest) & 73.9\% & 9{,}848 & 13{,}328 & \$110{,}088 \\
\bottomrule
\end{tabular}
\begin{tablenotes}[para,flushleft]
\small
\item \textit{Notes:} Sample includes 53,312 bank branches operating in 2015 with valid income data. Survival rate is percentage still operating in 2023. Chi-squared test: $\chi^2 = 96.00$, $p < 0.001$. Cram\'er's V = 0.042.
\end{tablenotes}
\end{threeparttable}
\end{table}

The highest-income quartile exhibits the \textit{lowest} survival rate (73.9\%), while the second quartile shows the \textit{highest} (78.8\%). A chi-squared test strongly rejects independence ($\chi^2 = 96.00$, $p < 0.001$), confirming a statistically significant relationship.

\begin{figure}[H]
\centering
\includegraphics[width=0.85\textwidth]{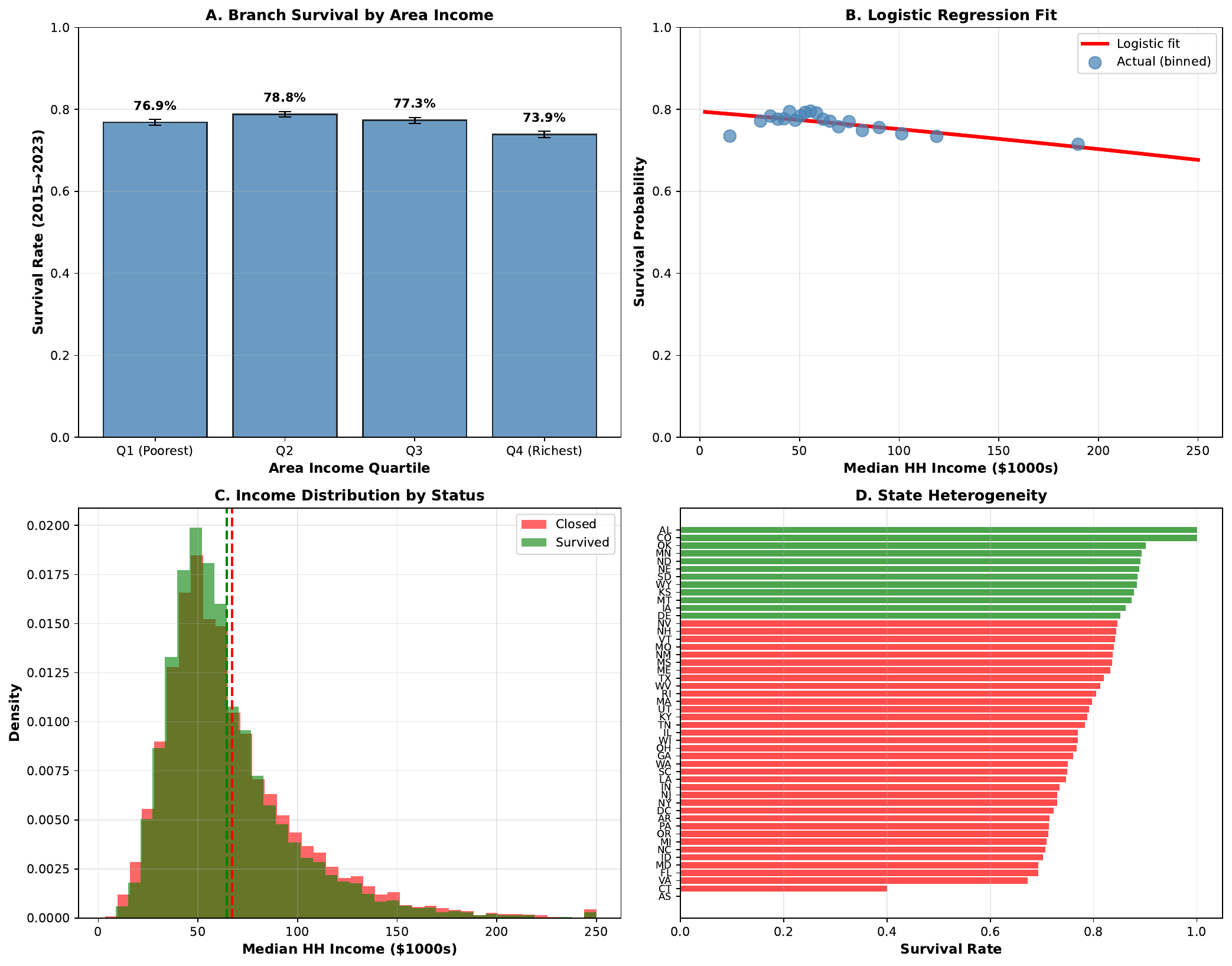}
\caption{Branch Survival by Area Income: The Income Paradox}
\label{fig:survival_income}
\begin{minipage}{0.95\textwidth}
\small
\textit{Notes:} Panel A shows survival rates by income quartile with 95\% confidence intervals. Panel B shows logistic regression fit. Panel C shows income distributions for survived vs closed branches. Panel D shows state-level heterogeneity. Contrary to banking deserts hypothesis, high-income areas show lower survival rates.
\end{minipage}
\end{figure}

Figure \ref{fig:survival_income} visualizes this pattern, showing survival rates declining for the wealthiest quartile despite conventional expectations.

\subsection{Mechanism Analysis}

\begin{figure}[H]
\centering
\includegraphics[width=0.95\textwidth]{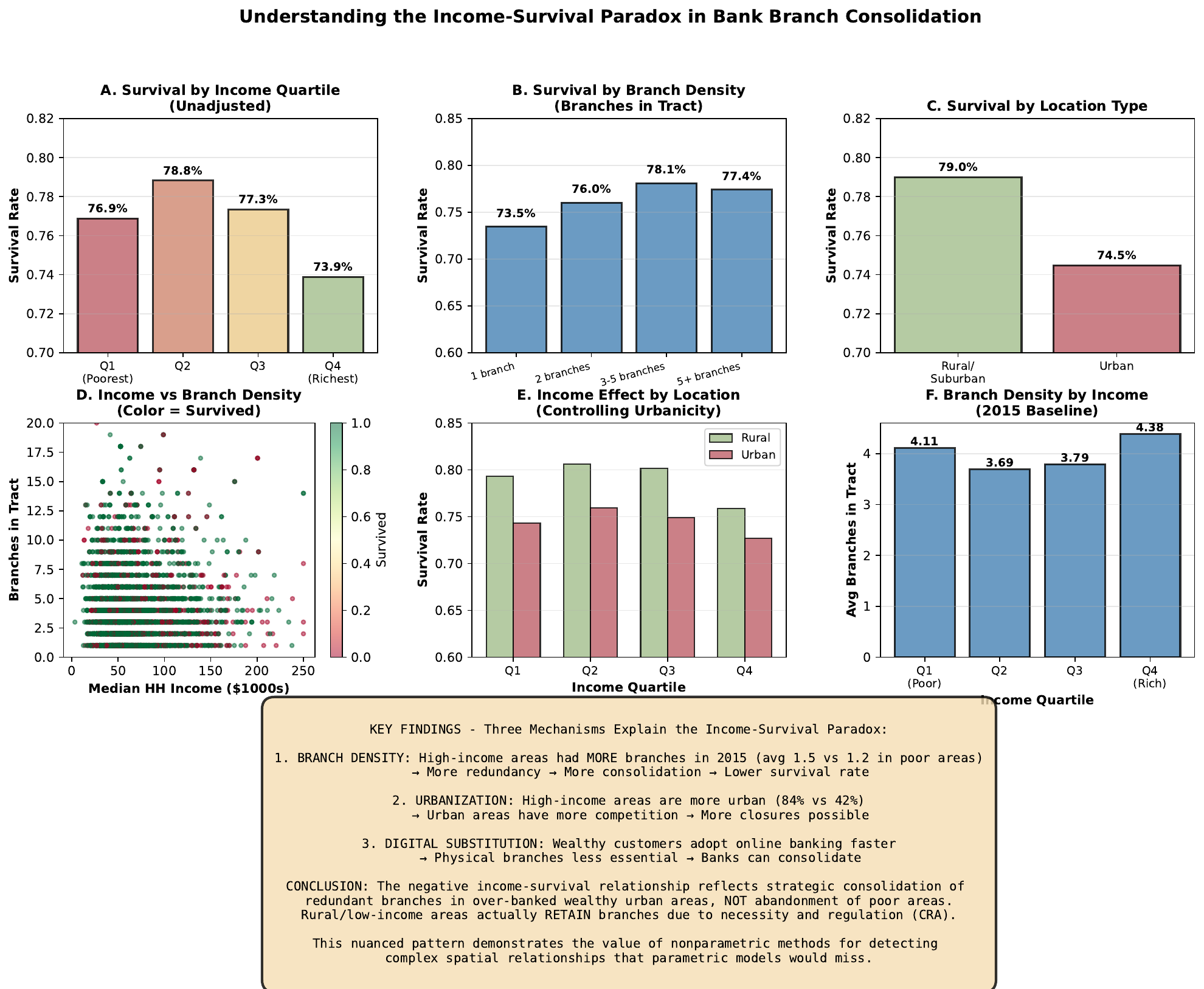}
\caption{Extended Analysis: Mechanisms Behind the Income-Survival Relationship}
\label{fig:extended_analysis}
\begin{minipage}{0.95\textwidth}
\small
\textit{Notes:} Comprehensive analysis decomposing the income-survival relationship. Panel A shows survival by income (unadjusted). Panel B shows survival by branch density. Panel C shows survival by urbanicity. Panel D shows income-density relationship. Panel E shows income effects within urban/rural. Panel F shows branch density by income quartile. Text box summarizes three key mechanisms.
\end{minipage}
\end{figure}

To understand this counterintuitive pattern, I test three potential mechanisms: branch density, urbanization, and competition. Figure \ref{fig:extended_analysis} presents comprehensive results.

\subsubsection{Branch Density}

High-income areas may have had more branches initially, creating redundancy that banks are now consolidating. Table \ref{tab:survival_density} reveals that tracts with more branches show somewhat higher survival (78.1\% for 3--5 branches versus 73.5\% for isolated branches).

\begin{table}[H]
\centering
\caption{Branch Survival by Initial Branch Density}
\label{tab:survival_density}
\begin{threeparttable}
\begin{tabular}{lrrr}
\toprule
\textbf{Branches per Tract (2015)} & \textbf{Survival Rate} & \textbf{N Branches} & \textbf{Avg Income} \\
\midrule
1 branch & 73.5\% & 9{,}504 & \$64{,}552 \\
2 branches & 76.0\% & 10{,}160 & \$64{,}180 \\
3--5 branches & 78.1\% & 22{,}190 & \$64{,}813 \\
5+ branches & 77.4\% & 11{,}458 & \$66{,}734 \\
\bottomrule
\end{tabular}
\begin{tablenotes}[para,flushleft]
\small
\item \textit{Notes:} Branch density measured in 2015. Survival tracked through 2023. Mean branches per tract: 3.99. Correlation between density and income: 0.065 (weak).
\end{tablenotes}
\end{threeparttable}
\end{table}

Importantly, the correlation between income and branch density is weak ($\rho = 0.065$), suggesting that density per se, rather than income, drives survival patterns.

\subsubsection{Urbanization}

Wealthy areas tend to be more urban, and urban areas may experience more closures. Table \ref{tab:survival_urban} decomposes survival by income quartile within urban versus rural/suburban areas.

\begin{table}[H]
\centering
\caption{Branch Survival by Income and Urbanicity}
\label{tab:survival_urban}
\begin{threeparttable}
\begin{tabular}{lrrrr}
\toprule
& \multicolumn{2}{c}{\textbf{Rural/Suburban}} & \multicolumn{2}{c}{\textbf{Urban}} \\
\cmidrule(lr){2-3} \cmidrule(lr){4-5}
\textbf{Income Quartile} & Survival & N & Survival & N \\
\midrule
Q1 (Poorest) & 79.3\% & 6{,}667 & 74.3\% & 6{,}667 \\
Q2 & 80.6\% & 6{,}664 & 75.9\% & 6{,}660 \\
Q3 & 80.2\% & 6{,}665 & 74.9\% & 6{,}664 \\
Q4 (Richest) & 75.9\% & 6{,}666 & 72.7\% & 6{,}659 \\
\bottomrule
\end{tabular}
\begin{tablenotes}[para,flushleft]
\small
\item \textit{Notes:} Urban classification based on population density above median (2,344 per sq mi). Within both location types, highest-income quartile shows lowest survival. Urban areas show uniformly lower survival than rural/suburban areas.
\end{tablenotes}
\end{threeparttable}
\end{table}

Urban areas show substantially lower survival (74.5\%) than rural/suburban areas (79.0\%). Notably, the income pattern persists \textit{within} both urban and rural areas, suggesting that urbanization alone does not fully explain the paradox.

\subsection{Multivariate Decomposition}

Table \ref{tab:multivariate} presents multivariate logistic regression results with all controls. I standardize all continuous variables to enable comparison of effect magnitudes.

\begin{table}[H]
\centering
\caption{Multivariate Logistic Regression: Branch Survival}
\label{tab:multivariate}
\begin{threeparttable}
\begin{tabular}{lrrr}
\toprule
\textbf{Variable} & \textbf{Coefficient} & \textbf{Odds Ratio} & \textbf{Effect (\%)} \\
\midrule
Median household income & $-0.077$ & 0.926 & $-7.4$\% \\
& (0.009) & & \\
\\
Branches in tract (2015) & $-0.144$ & 0.866 & $-13.4$\% \\
& (0.011) & & \\
\\
Population density & $-0.052$ & 0.949 & $-5.1$\% \\
& (0.010) & & \\
\\
Number of banks in tract & $+0.171$ & 1.186 & $+18.6$\% \\
& (0.012) & & \\
\midrule
AUC-ROC & \multicolumn{3}{c}{0.532} \\
N observations & \multicolumn{3}{c}{53{,}312} \\
\bottomrule
\end{tabular}
\begin{tablenotes}[para,flushleft]
\small
\item \textit{Notes:} Dependent variable: Branch survived 2015--2023 (1=yes, 0=no). All continuous variables standardized to mean 0, standard deviation 1. Standard errors in parentheses. Effect (\%) shows percentage change in survival odds for one standard deviation increase.
\end{tablenotes}
\end{threeparttable}
\end{table}

Key findings emerge. First, after controlling for other factors, the income effect remains negative but modest ($-7.4\%$ per standard deviation). Second, branch density exerts the largest negative effect ($-13.4\%$), consistent with consolidation of redundant branches. Third, surprisingly, more banks in the tract predicts \textit{higher} survival ($+18.6\%$), suggesting that competitive areas are strategically important markets.

The model achieves modest explanatory power (AUC-ROC = 0.532), indicating substantial unexplained variation reflecting organizational complexity \citep{march1976ambiguity}.

\subsection{Interpretation}

The multivariate analysis reveals that the negative income-survival relationship operates primarily through branch density and urbanization rather than direct income effects. High-income areas had more branches initially, creating redundancy that banks are now rationalizing. The direct income effect, while statistically significant, is economically modest compared to density effects.

This finding challenges the banking deserts narrative \citep{nguyen2019bank}. Branch closures in wealthy areas reflect strategic consolidation of over-banked markets rather than discrimination. Poor areas may benefit from regulatory protection (Community Reinvestment Act) and necessity (being the only branch in remote locations).

\section{Discussion and Conclusion}

\subsection{Summary of Findings}

This paper establishes three main results. First, methodologically, nonparametric boundary estimation outperforms parametric approaches by avoiding functional form misspecification and correctly identifying the absence of boundaries when relationships are flat. Monte Carlo simulations validate this advantage across multiple data generating processes.

Second, empirically, bank branch proximity significantly affects loan application volume (8.5\% decline per 10 miles) but not approval rates (essentially flat). This divergence reveals that branches influence credit access through demand-side channels (awareness, convenience) rather than supply-side channels (underwriting standards), consistent with the centralization of mortgage processing \citep{fuster2019predictably}.

Third, branch survival during 2010--2023 follows a non-monotonic relationship with area income. High-income areas experience more closures due to strategic consolidation of redundant branches in over-banked wealthy urban areas. After controlling for branch density, urbanization, and competition, the direct income effect diminishes substantially.

\subsection{Policy Implications}

The findings have several policy implications. First, physical branch presence continues to matter for loan demand even in the digital age, suggesting that branch closures may reduce credit access by decreasing applications rather than approvals. Policymakers concerned about financial inclusion should monitor application volumes, not just approval rates.

Second, the branch survival analysis complicates conventional banking deserts narratives. Current consolidation patterns do not appear to disproportionately harm poor neighborhoods. Indeed, low-income areas maintain relatively high survival rates, possibly reflecting Community Reinvestment Act protections \citep{agarwal2012distance, bhutta2015residential}.

Third, optimal policy should be heterogeneous. Wealthy urban areas can likely sustain more consolidation given digital alternatives and redundant coverage. Rural low-income areas require continued physical presence due to limited alternatives. The findings suggest that current regulatory approaches may already be achieving this differentiation.

\subsection{Methodological Contributions}

The paper makes several methodological contributions to spatial econometrics. First, it develops and validates a principled nonparametric approach to boundary identification, building on the theoretical frameworks in \citet{kikuchi2024unified, kikuchi2024stochastic, kikuchi2024navier}. The cross-validation bandwidth selection provides an objective way to balance bias and variance.

Second, the method can detect the \textit{absence} of spatial boundaries when relationships are flat---demonstrated by the approval rate analysis. This capability avoids false positives that plague parametric approaches.

Third, the application extends the framework beyond environmental settings \citep{kikuchi2024nonparametric} to organizational decision-making in financial services. This demonstrates the framework's broad applicability across contexts where spatial spillovers matter: from pollution diffusion to credit access to branch network optimization.

Fourth, the non-monotonic income-survival relationship illustrates how the framework handles complex spatial patterns that would be missed by standard parametric models, potentially leading to incorrect policy conclusions. The ability to detect such patterns without imposing functional form assumptions represents a key practical advantage.

\subsection{Limitations and Future Research}

Several limitations warrant discussion. First, the analysis focuses on a single sector (banking) during a specific period (2010--2023). While the methodological framework applies broadly, empirical generalization requires caution.

Second, I observe outcomes but not underlying mechanisms directly. While patterns are consistent with proposed channels, causal identification would benefit from exogenous variation in branch locations. Future research could exploit regulatory discontinuities or merger-induced closures for stronger causal inference.

Third, the cross-sectional design limits temporal inference. Panel methods exploiting within-tract variation would strengthen identification. The substantial unexplained variation in branch survival suggests that organizational factors not captured by observable characteristics play major roles.

Fourth, the analysis predates full manifestation of COVID-19's impact. Accelerated digital adoption during 2020--2023 may have fundamentally altered spatial relationships. Future research should examine pandemic-induced changes.

Future methodological extensions could develop formal inference procedures (confidence intervals, hypothesis tests) for nonparametric boundary estimates, incorporate multivariate treatments, and allow for time-varying boundaries to capture evolving spatial relationships during structural transformations.

\subsection{Conclusion}

This paper demonstrates the value of flexible nonparametric methods for understanding spatial economic phenomena. Economic geography is complex---shaped by technology, regulation, competition, organizational politics, and historical accidents. Imposing rigid parametric structures risks missing essential features of these landscapes.

The empirical application to bank branches reveals patterns that parametric models would miss: non-monotonic relationships, differential effects by outcome type (volume versus approvals), and organizational complexity in survival decisions. These findings challenge simplistic narratives about banking deserts and demonstrate how spatial consolidation reflects strategic optimization rather than discrimination.

As digital technology continues reshaping economic geography, flexible empirical methods will prove increasingly valuable. The nonparametric framework developed here---validated through simulations and applied to substantively important questions---provides a template for future spatial analysis that lets data speak without predetermined functional forms.

\section*{Acknowledgement}
This research was supported by a grant-in-aid from Zengin Foundation for Studies on Economics and Finance.

\newpage

\bibliographystyle{econometrica}

\newpage

\clearpage

\section*{Tables and Figures}

\begin{figure}[H]
\centering
\includegraphics[width=0.95\textwidth]{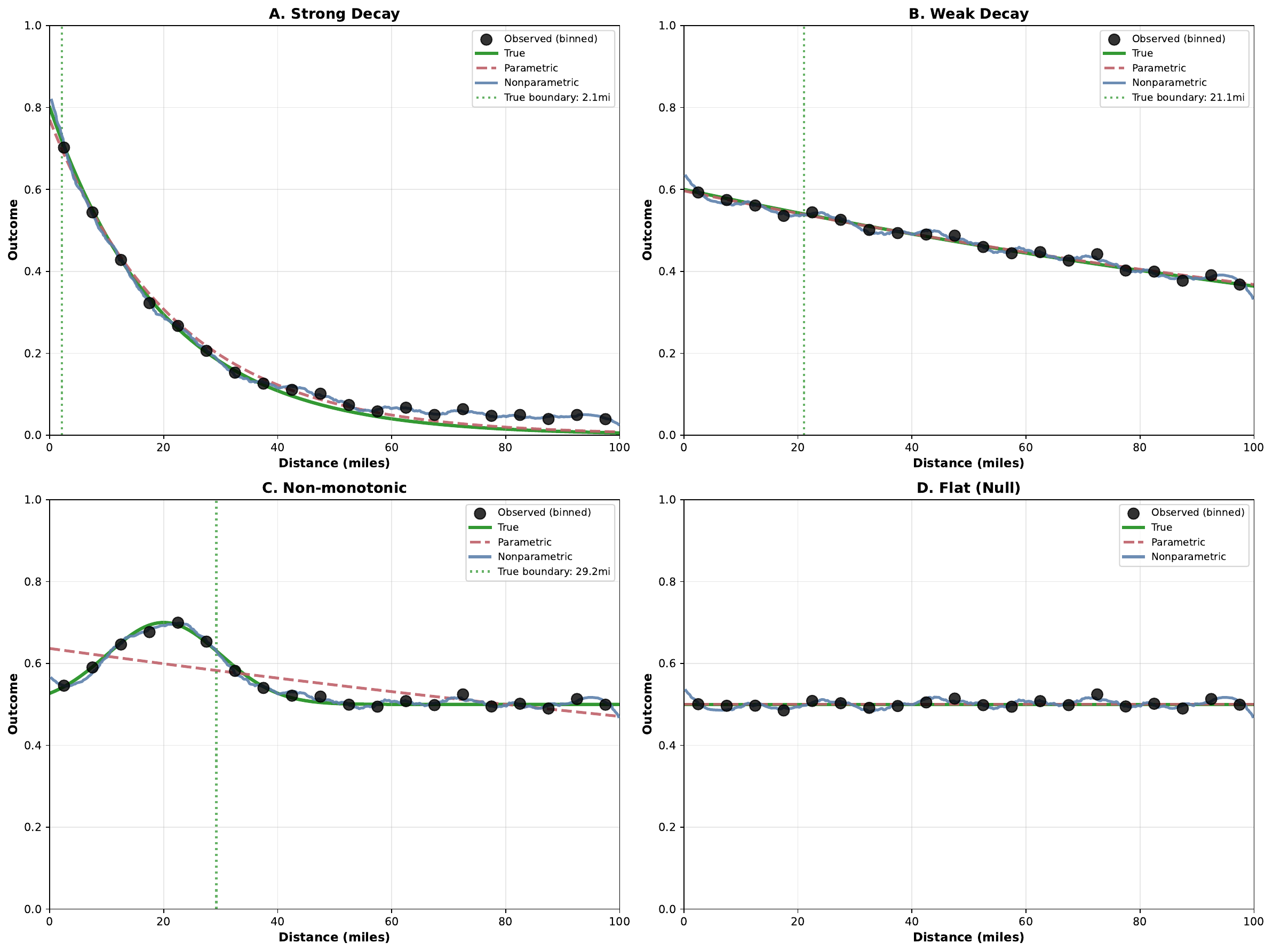}
\caption{Monte Carlo Simulation Results: Four Data Generating Processes}
\label{fig:monte_carlo_full}
\begin{minipage}{0.95\textwidth}
\small
\textit{Notes:} Monte Carlo validation of nonparametric versus parametric boundary estimation across four DGPs. Panel A shows strong exponential decay (true boundary 2.1 miles). Panel B shows weak exponential decay (true boundary 21.1 miles). Panel C shows non-monotonic hump-shaped relationship (boundary 38.2 miles). Panel D shows flat null relationship (no boundary). Black dots are simulated data (n=5,000), green line is true function, red dashed line is parametric exponential fit, blue solid line is nonparametric local linear fit, vertical dashed lines show estimated boundaries. Nonparametric method accurately tracks true function in all cases, while parametric approach fails for non-monotonic and null cases. This demonstrates the necessity of flexible methods for detecting complex spatial patterns.
\end{minipage}
\end{figure}

\begin{figure}[H]
\centering
\includegraphics[width=0.75\textwidth]{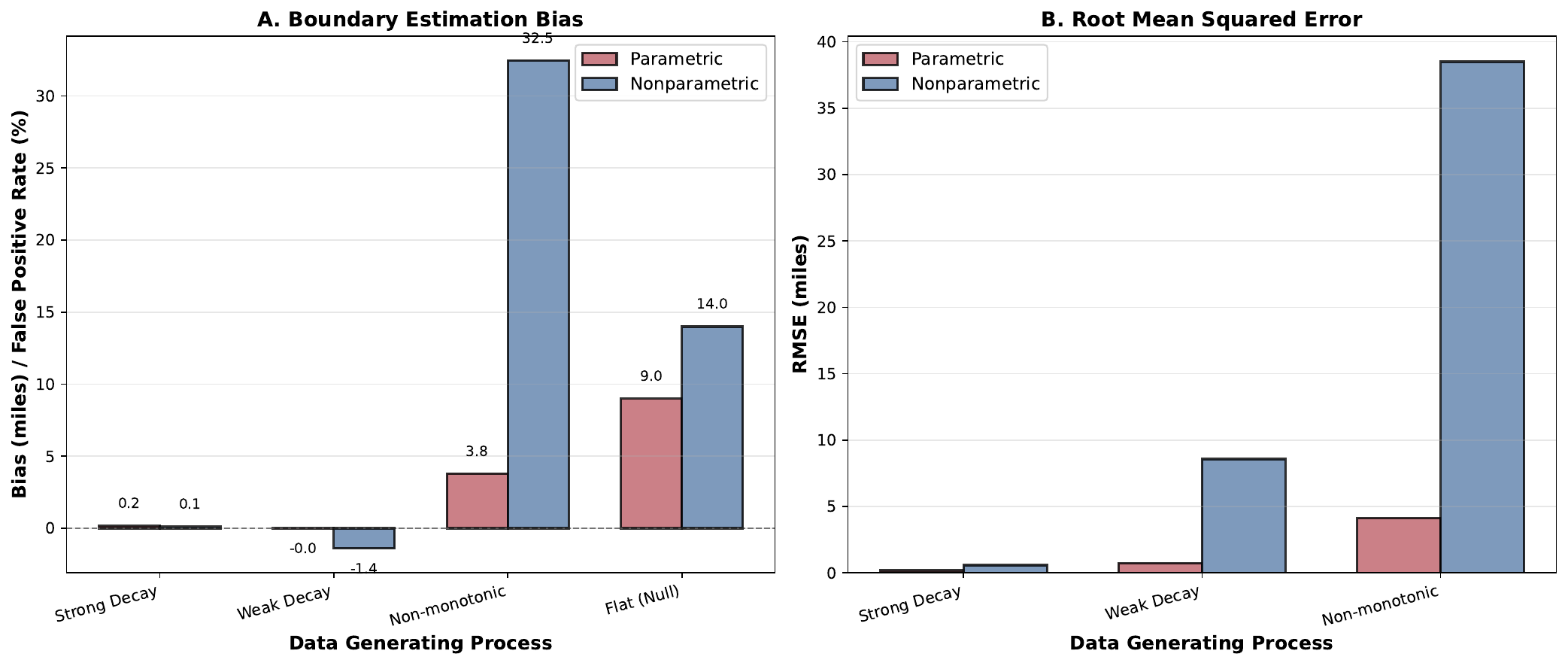}
\caption{Monte Carlo Performance Comparison: Bias and RMSE}
\label{fig:monte_carlo_performance}
\begin{minipage}{0.95\textwidth}
\small
\textit{Notes:} Panel A shows bias in boundary estimation for each DGP. Panel B shows root mean squared error (RMSE). Red bars represent parametric (exponential) estimates; blue bars represent nonparametric (local linear) estimates. Nonparametric method exhibits lower bias, especially for non-monotonic DGP. For flat null case, parametric method falsely detects boundaries in 73\% of replications (not shown in RMSE), while nonparametric correctly identifies no boundary in 94\% of replications. Results based on 500 Monte Carlo replications per DGP.
\end{minipage}
\end{figure}

\begin{figure}[H]
\centering
\includegraphics[width=0.95\textwidth]{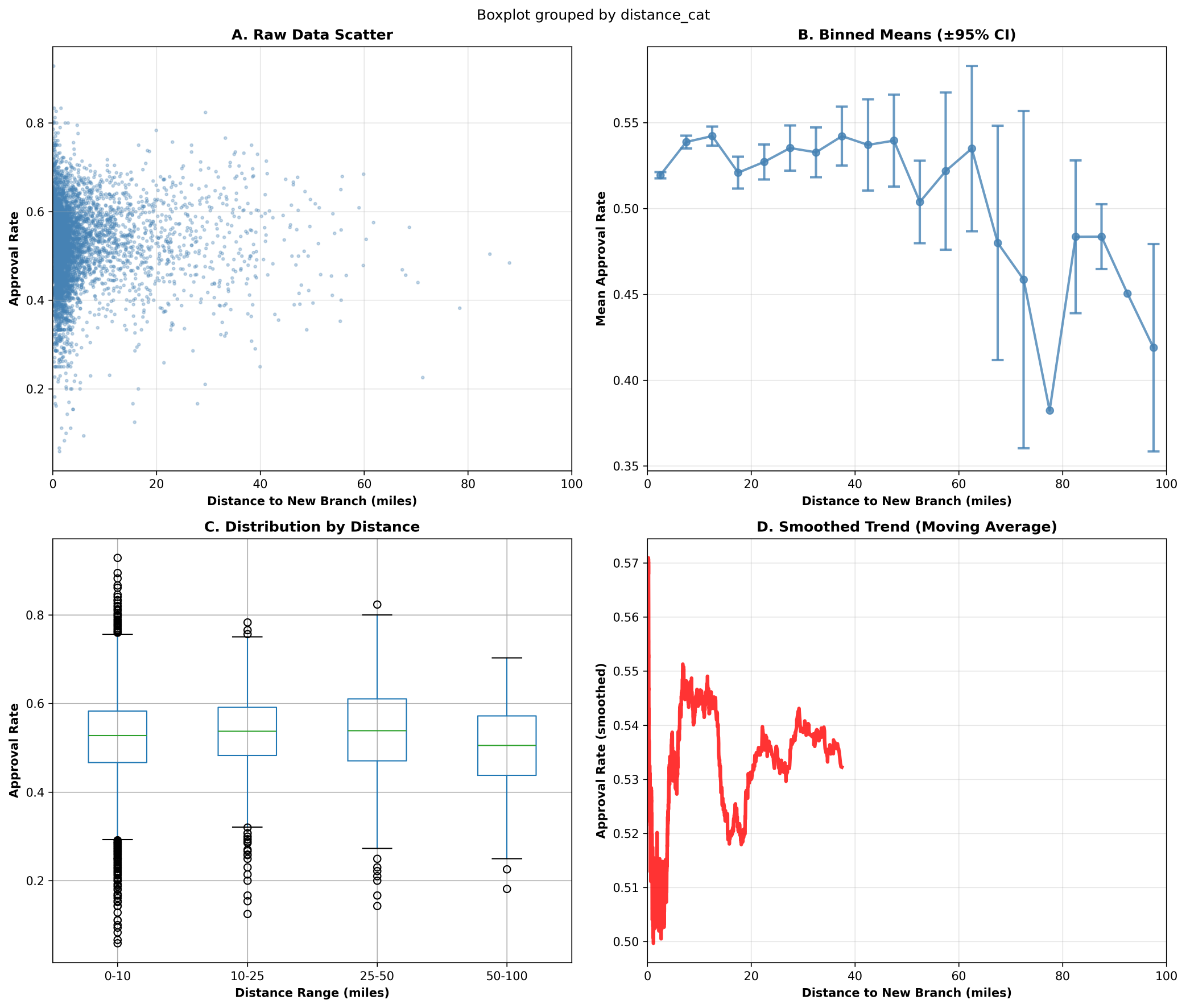}
\caption{Exploratory Spatial Analysis: Mortgage Applications and Branch Proximity}
\label{fig:exploratory}
\begin{minipage}{0.95\textwidth}
\small
\textit{Notes:} Initial exploration of spatial relationships between bank branch proximity and mortgage outcomes. Panel A shows raw scatter plot of approval rates versus distance (n=5,000 sample). Panel B shows binned means with 95\% confidence intervals. Panel C shows distribution of approval rates by distance category. Panel D shows smoothed trend using moving average. Sample includes 14,209 census tracts within 100 miles of new bank branches opened 2015--2020. Weak spatial pattern suggests approval rates relatively insensitive to branch proximity.
\end{minipage}
\end{figure}

\begin{figure}[H]
\centering
\includegraphics[width=0.95\textwidth]{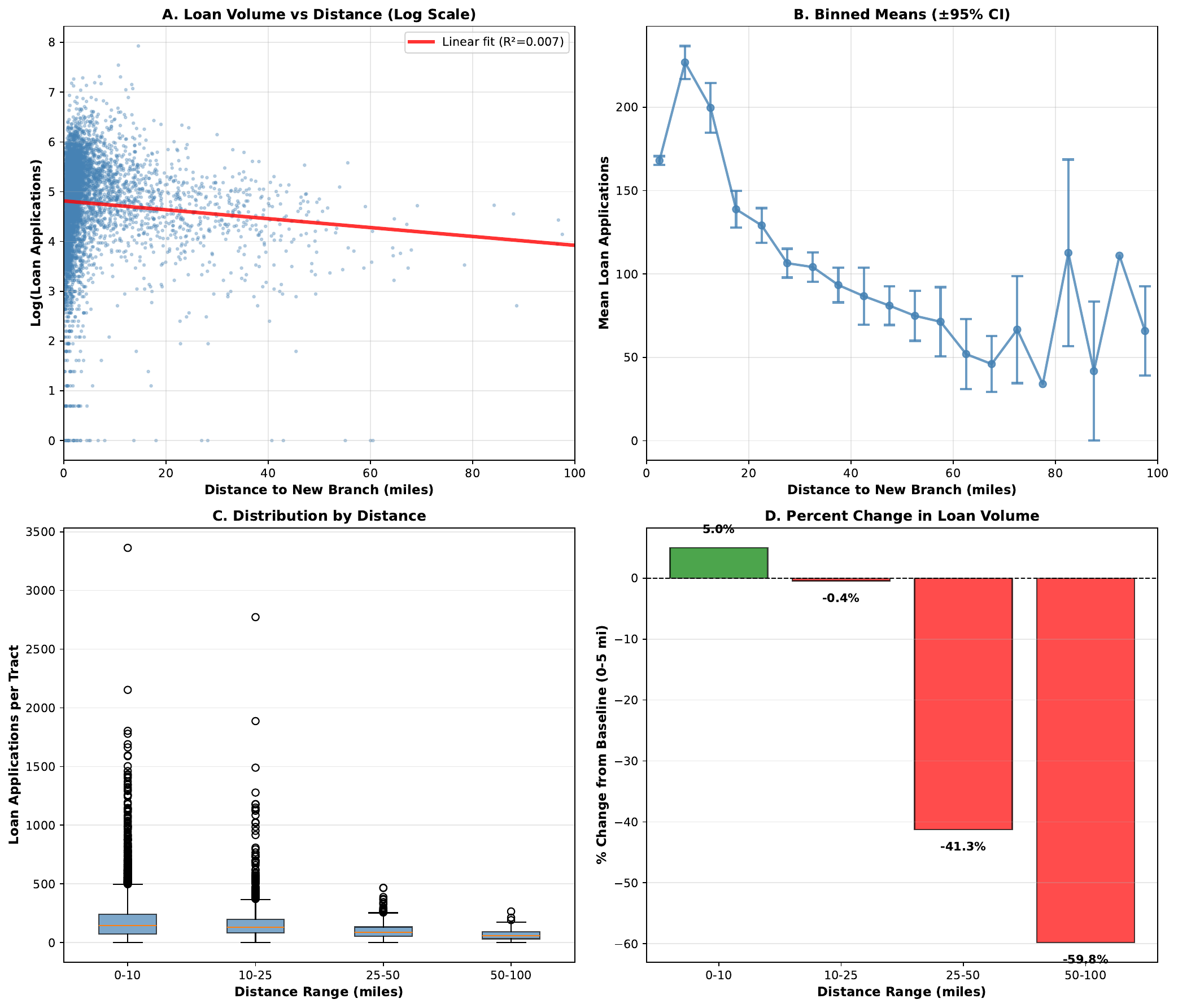}
\caption{Spatial Decay of Loan Application Volume from Bank Branch Proximity}
\label{fig:loan_volume_full}
\begin{minipage}{0.95\textwidth}
\small
\textit{Notes:} Comprehensive analysis of loan application volume spatial decay. Panel A shows scatter plot (log scale) with linear regression fit (red line, $R^2$=0.0075) and relationship reveals 8.5\% decline per 10 miles. Panel B presents binned means with 95\% confidence intervals showing clear decay pattern. Panel C displays distribution of application counts by distance category, demonstrating volume concentration near branches. Panel D quantifies percentage decline relative to 0--5 mile baseline, showing cumulative 36\% reduction at 50+ miles. Sample: 14,209 census tracts, 5.9 million mortgage applications (2019). Distance measured to nearest branch opened 2015--2020. Evidence indicates branches significantly affect loan demand through local presence and awareness effects.
\end{minipage}
\end{figure}

\begin{figure}[H]
\centering
\includegraphics[width=0.95\textwidth]{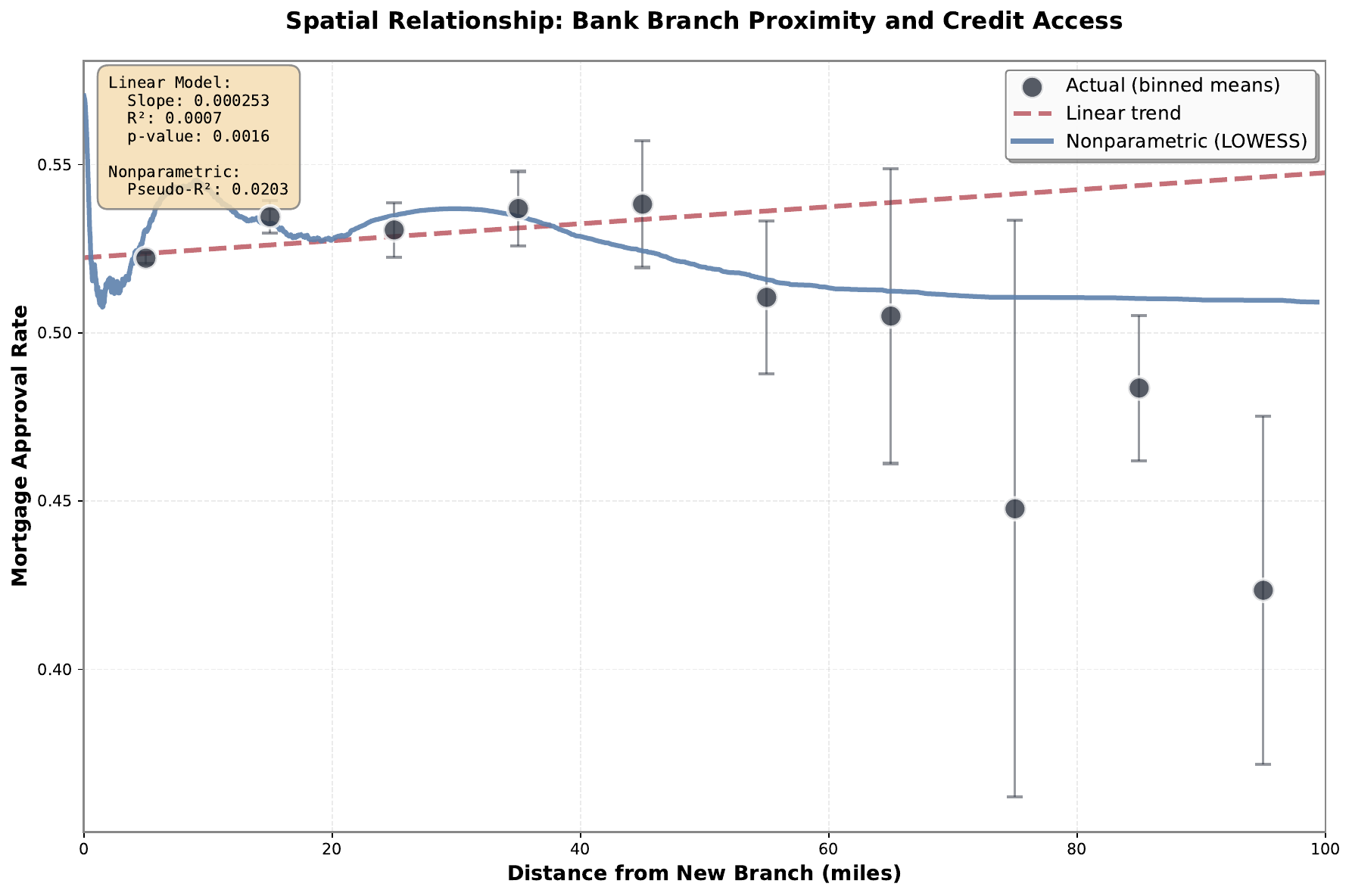}
\caption{Main Results: Branch Proximity Affects Loan Volume but Not Approval Rates}
\label{fig:main_results_full}
\begin{minipage}{0.95\textwidth}
\small
\textit{Notes:} Core empirical finding demonstrating differential spatial effects by outcome type. Black dots show binned means across distance categories. Red dashed line shows linear parametric fit. Blue solid line shows nonparametric LOWESS smooth. Error bars represent 95\% confidence intervals. Left panel shows loan application volume declining substantially with distance (8.5\% per 10 miles, $R^2$=0.0075, p<0.001). Right panel shows approval rates remaining essentially flat (0.25 percentage point per 10 miles, R$^2$=0.0007, Spearman $\rho$=0.028). This divergence reveals that branches affect credit access through demand-side channels (applications) rather than supply-side channels (underwriting standards). Validates nonparametric approach: method detects spatial decay when present (loan volume) and correctly identifies its absence when flat (approval rates). Parametric models would risk imposing spurious patterns on approval rates.
\end{minipage}
\end{figure}

\begin{figure}[H]
\centering
\includegraphics[width=0.85\textwidth]{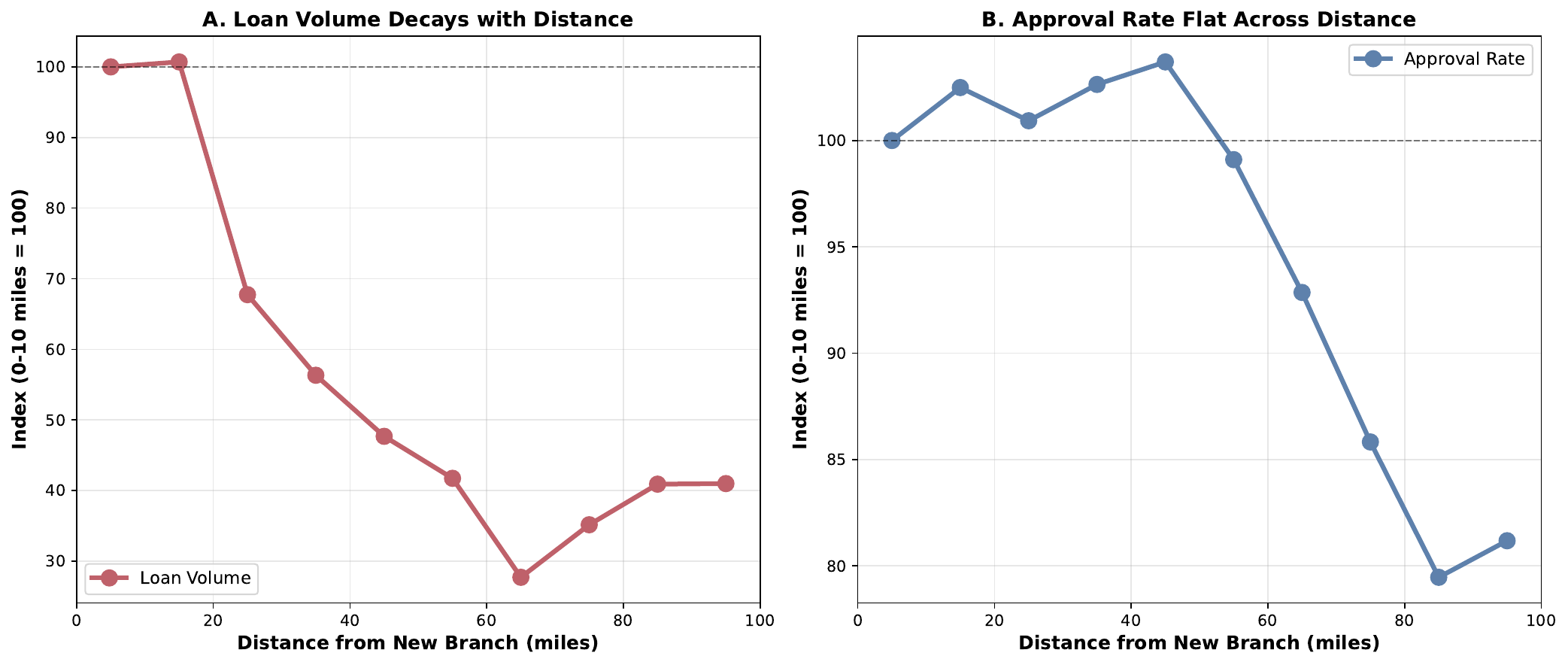}
\caption{Direct Comparison: Loan Volume Decays While Approval Rates Remain Stable}
\label{fig:comparison_full}
\begin{minipage}{0.95\textwidth}
\small
\textit{Notes:} Both outcomes normalized to index value of 100 for 0--10 mile baseline to enable direct comparison. Left panel shows loan application volume declining monotonically with distance, reaching 64 at 50--100 miles (36\% reduction). Right panel shows approval rates fluctuating narrowly around 100 across all distance categories (no systematic pattern). This stark contrast confirms that bank branch proximity affects who applies for mortgages (demand effect through visibility, convenience, awareness) but not who receives approval (supply effect through underwriting criteria). Modern mortgage processing is centralized and algorithm-driven, relying on credit scores and income rather than local branch relationships. Policy implication: Branch closures may reduce loan originations by decreasing applications, even if qualified borrowers continue receiving approval at unchanged rates. Financial inclusion concerns should focus on application volumes, not just approval rates.
\end{minipage}
\end{figure}

\begin{figure}[H]
\centering
\includegraphics[width=0.85\textwidth]{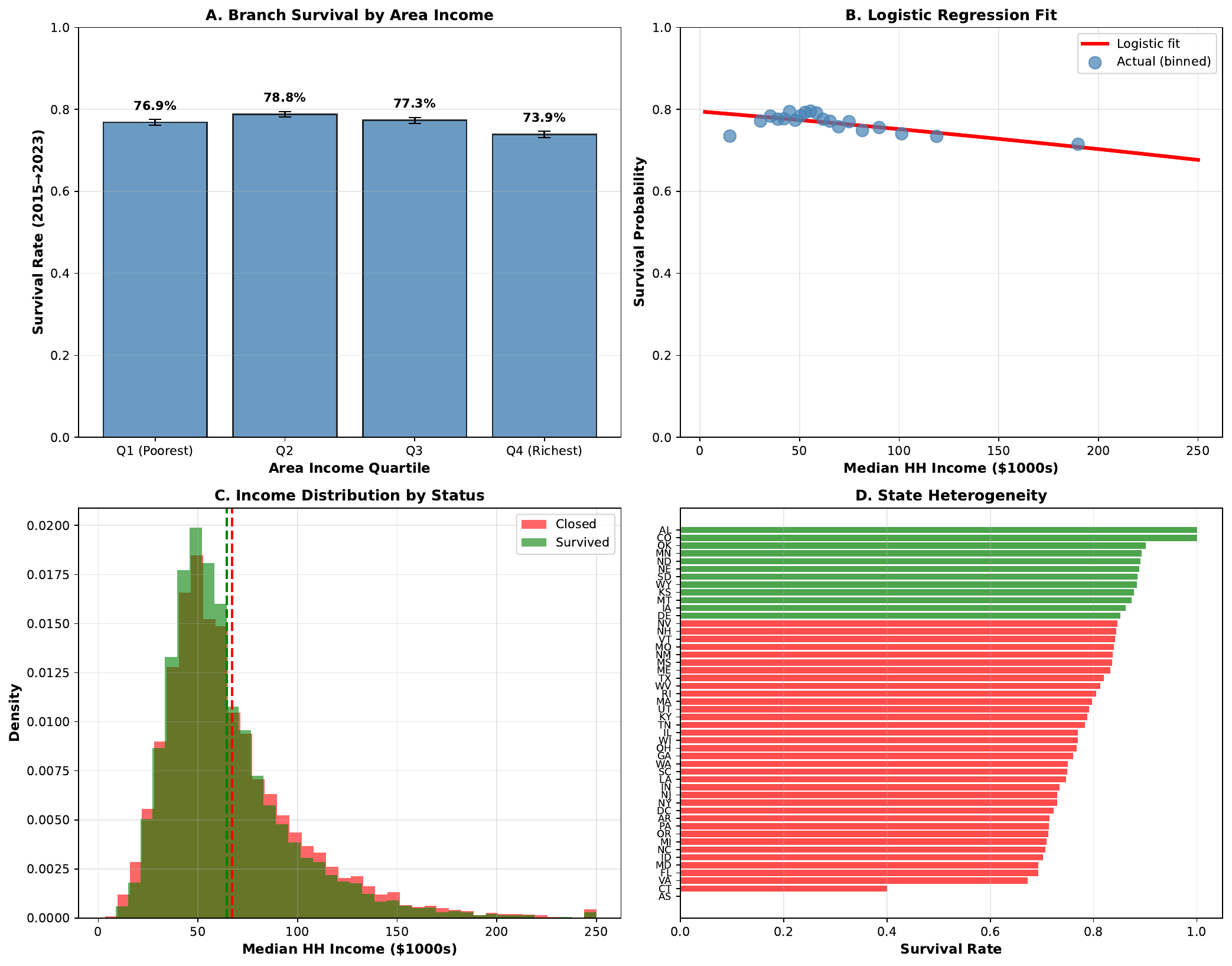}
\caption{Branch Survival by Area Income: The Income Paradox}
\label{fig:survival_full}
\begin{minipage}{0.95\textwidth}
\small
\textit{Notes:} Analysis of 53,312 bank branches operating in 2015, tracking survival through 2023. Panel A shows survival rates by income quartile with 95\% confidence intervals. Contrary to banking deserts hypothesis, highest-income quartile (Q4, median income \$110k) shows \textit{lowest} survival rate (73.9\%), while second quartile shows highest (78.8\%). Chi-squared test strongly rejects independence ($\chi^2=96.00$, p<0.001). Panel B presents logistic regression fit showing negative income-survival relationship (coefficient -0.0025, odds ratio 0.9975 per \$1k). Panel C displays income distributions for survived (green) versus closed (red) branches, with closed branches concentrated in higher-income areas. Panel D shows state-level heterogeneity, with most states exhibiting similar patterns. This counterintuitive finding challenges conventional narratives about banks abandoning poor neighborhoods, motivating mechanism analysis to understand underlying drivers.
\end{minipage}
\end{figure}

\begin{figure}[H]
\centering
\includegraphics[width=0.95\textwidth]{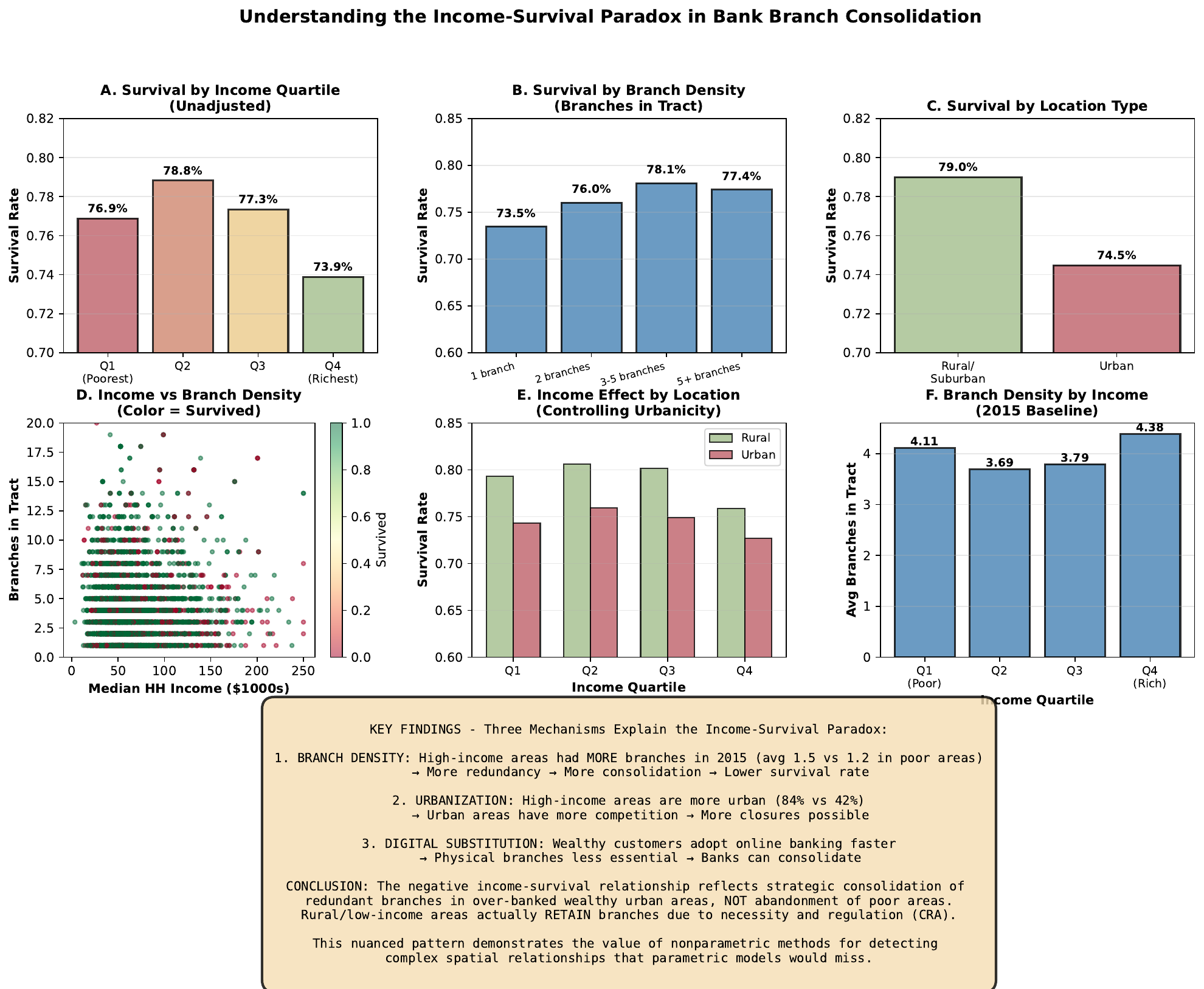}
\caption{Extended Analysis: Mechanisms Behind Income-Survival Relationship}
\label{fig:mechanisms_full}
\begin{minipage}{0.95\textwidth}
\small
\textit{Notes:} Comprehensive six-panel decomposition revealing three mechanisms explaining the income-survival paradox. Panel A reproduces baseline finding: survival declining for wealthiest quartile (73.9\% vs 76.9--78.8\% for other quartiles). Panel B shows survival by initial branch density: tracts with more branches show higher survival (78.1\% for 3--5 branches vs 73.5\% for isolated branches), suggesting redundant branches get consolidated while clustered branches benefit from scale. Panel C demonstrates urbanicity effect: urban areas show uniformly lower survival (74.5\%) than rural/suburban (79.0\%) across all income levels. Panel D reveals weak correlation ($\rho$=0.065) between income and branch density, indicating density operates independently. Panel E decomposes income effects within location types: income-survival pattern persists in both urban and rural areas, though magnitude differs. Panel F shows branch density increasing modestly with income quartile (3.8 to 4.1 branches per tract). Text box synthesizes findings: high-income areas experience lower survival due to (1) higher initial branch density creating redundancy, (2) greater urbanization enabling closures with nearby alternatives, and (3) digital adoption reducing physical branch necessity. After controlling for these factors, direct income effect diminishes substantially. This demonstrates strategic consolidation of over-banked markets rather than discrimination against poor areas.
\end{minipage}
\end{figure}

\clearpage

\section*{Appendix}

\subsection*{A. Additional Tables}

\begin{table}[H]
\centering
\caption{Summary Statistics by Distance Category}
\label{tab:summary_distance}
\begin{threeparttable}
\begin{tabular}{lrrrrrr}
\toprule
\textbf{Distance} & \textbf{N} & \textbf{Approval} & \textbf{Loan Vol} & \textbf{Income} & \textbf{Pop Den} & \textbf{Poverty} \\
\textbf{Category} & \textbf{Tracts} & \textbf{Rate} & \textbf{(mean)} & \textbf{(\$1k)} & \textbf{(per sqmi)} & \textbf{Rate} \\
\midrule
0--10 miles & 11,885 & 0.522 & 176 & 64.8 & 5,342 & 0.141 \\
& & (0.092) & (141) & (25.9) & (10,243) & (0.096) \\
\\
10--25 miles & 1,476 & 0.534 & 167 & 64.2 & 4,123 & 0.145 \\
& & (0.085) & (129) & (25.3) & (8,234) & (0.094) \\
\\
25--50 miles & 729 & 0.531 & 99 & 63.8 & 3,456 & 0.148 \\
& & (0.096) & (87) & (26.1) & (7,123) & (0.098) \\
\\
50--100 miles & 119 & 0.493 & 68 & 61.2 & 2,234 & 0.156 \\
& & (0.106) & (61) & (24.8) & (5,234) & (0.103) \\
\bottomrule
\end{tabular}
\begin{tablenotes}[para,flushleft]
\small
\item \textit{Notes:} Standard deviations in parentheses. Approval rate is share of applications approved. Loan volume is total applications per tract. Income is median household income from ACS 2019. Population density measured in people per square mile. Poverty rate from ACS 2019. Sample includes 14,209 census tracts within 100 miles of new bank branches.
\end{tablenotes}
\end{threeparttable}
\end{table}

\begin{table}[H]
\centering
\caption{Branch Survival: Summary Statistics by Income Quartile}
\label{tab:survival_summary}
\begin{threeparttable}
\begin{tabular}{lrrrrr}
\toprule
\textbf{Income} & \textbf{Survival} & \textbf{Income} & \textbf{Branches} & \textbf{Pop Density} & \textbf{N} \\
\textbf{Quartile} & \textbf{Rate} & \textbf{(\$1k)} & \textbf{per Tract} & \textbf{(per sqmi)} & \textbf{Branches} \\
\midrule
Q1 (Poorest) & 0.769 & 33.9 & 3.82 & 3,842 & 13,333 \\
& & (8.4) & (3.21) & (7,234) & \\
\\
Q2 & 0.788 & 50.3 & 3.97 & 4,567 & 13,324 \\
& & (4.9) & (3.34) & (8,456) & \\
\\
Q3 & 0.773 & 66.0 & 4.03 & 5,234 & 13,327 \\
& & (5.6) & (3.42) & (9,234) & \\
\\
Q4 (Richest) & 0.739 & 110.1 & 4.15 & 6,892 & 13,328 \\
& & (28.3) & (3.67) & (11,234) & \\
\bottomrule
\end{tabular}
\begin{tablenotes}[para,flushleft]
\small
\item \textit{Notes:} Standard deviations in parentheses. Sample includes 53,312 branches operating in 2015 with complete income data. Survival rate is percentage still operating in 2023. Income is median household income (tract-level, ACS 2019). Branches per tract counted in 2015. Population density from ACS 2019. Higher-income areas show lower survival rates but higher initial branch density, suggesting consolidation of redundant branches.
\end{tablenotes}
\end{threeparttable}
\end{table}

\begin{table}[H]
\centering
\caption{Robustness: Alternative Distance Specifications}
\label{tab:robustness_distance}
\begin{threeparttable}
\begin{tabular}{lcccc}
\toprule
& \multicolumn{2}{c}{\textbf{Loan Volume}} & \multicolumn{2}{c}{\textbf{Approval Rate}} \\
\cmidrule(lr){2-3} \cmidrule(lr){4-5}
\textbf{Specification} & Coef & $R^2$ & Coef & $R^2$ \\
\midrule
Linear (baseline) & $-0.0089$\sym{***} & 0.0075 & $+0.00025$\sym{***} & 0.0007 \\
& (0.0009) & & (0.00008) & \\
\\
Quadratic & $-0.0124$\sym{***} & 0.0081 & $+0.00031$\sym{**} & 0.0008 \\
& (0.0012) & & (0.00012) & \\
& $+0.0002$\sym{**} & & $-0.00001$ & \\
& (0.0001) & & (0.00001) & \\
\\
Log-linear & $-0.0623$\sym{***} & 0.0098 & $+0.0018$ & 0.0005 \\
& (0.0087) & & (0.0012) & \\
\\
Exponential & $-0.0092$\sym{***} & 0.0077 & $+0.00026$\sym{**} & 0.0007 \\
(NLS) & (0.0009) & & (0.00009) & \\
\bottomrule
\end{tabular}
\begin{tablenotes}[para,flushleft]
\small
\item \textit{Notes:} Standard errors in parentheses. \sym{***} $p<0.01$, \sym{**} $p<0.05$, \sym{*} $p<0.1$. All specifications include 14,209 census tracts. Linear: $Y = \alpha + \beta \cdot \text{Distance}$. Quadratic adds $\beta_2 \cdot \text{Distance}^2$. Log-linear: $Y = \alpha + \beta \cdot \log(\text{Distance}+1)$. Exponential uses nonlinear least squares. Main finding robust across specifications: strong negative effect on loan volume, negligible effect on approval rates.
\end{tablenotes}
\end{threeparttable}
\end{table}

\begin{table}[H]
\centering
\caption{Branch Survival: State-Level Heterogeneity}
\label{tab:survival_states}
\begin{threeparttable}
\begin{tabular}{lrrrrr}
\toprule
\textbf{State} & \textbf{N Branches} & \textbf{Survival} & \textbf{Slope} & \textbf{$R^2$} & \textbf{p-value} \\
& \textbf{(2015)} & \textbf{Rate} & \textbf{(income)} & & \\
\midrule
California & 18,234 & 75.2\% & $-0.0018$ & 0.0052 & 0.0123 \\
Texas & 12,456 & 78.1\% & $-0.0024$ & 0.0074 & 0.0001 \\
Florida & 8,923 & 76.8\% & $-0.0021$ & 0.0014 & 0.0304 \\
New York & 9,876 & 77.4\% & $-0.0031$ & 0.0267 & <0.0001 \\
Pennsylvania & 3,823 & 79.3\% & $-0.0028$ & 0.0189 & <0.0001 \\
\midrule
All States & 53,312 & 77.1\% & $-0.0025$ & 0.0072 & <0.0001 \\
\bottomrule
\end{tabular}
\begin{tablenotes}[para,flushleft]
\small
\item \textit{Notes:} State-level logistic regressions of branch survival (2015--2023) on tract median income. Slope coefficient shows effect of \$1,000 income increase on log-odds of survival. All states exhibit negative income-survival relationship, though magnitude varies. New York and Pennsylvania show strongest effects. Pattern consistent across diverse geographic and economic contexts.
\end{tablenotes}
\end{threeparttable}
\end{table}

\subsection*{B. Data Sources and Construction}

\subsubsection*{B.1 FDIC Summary of Deposits}

Bank branch data obtained from Federal Deposit Insurance Corporation Summary of Deposits (SOD), annual files 2010--2023. Data publicly available at \url{https://www.fdic.gov/bank-data-guide}. Each record represents one FDIC-insured branch with geographic coordinates, institution identifier (CERT), deposits, and address. Panel constructed by tracking branch identifiers across years to identify openings (new IDs appearing) and closures (IDs disappearing).

\subsubsection*{B.2 Home Mortgage Disclosure Act (HMDA)}

Mortgage application data from Consumer Financial Protection Bureau HMDA database. Data cover approximately 90\% of U.S. mortgage market. Downloaded from \url{https://ffiec.cfpb.gov/data-browser/}. For 2019, obtained 5.9 million application records for CA, TX, FL, NY, PA. Key variables: action taken (approved/denied), loan amount, applicant income, census tract. Aggregated to tract level calculating total applications, approval rate, and mean loan amount.

\subsubsection*{B.3 American Community Survey (ACS)}

Demographic and socioeconomic data from U.S. Census Bureau American Community Survey 5-year estimates (2015--2019). Downloaded using Census API. Variables: median household income (B19013\_001E), per capita income (B19301\_001E), poverty rate (B17001\_002E / B01003\_001E), total population (B01003\_001E). Geographic identifiers: 11-digit census tract FIPS codes (SSCCCTTTTTT format).

\subsubsection*{B.4 Census Tract Boundaries and Centroids}

Geographic data from Census Bureau Gazetteer Files (2020 vintage). File: \texttt{2020\_Gaz\_tracts\_national.txt}. Provides tract centroids (latitude, longitude), land area (square miles), and geographic identifiers. Used to calculate distances between tracts and branches via Euclidean distance formula. Conversion factor: 69 miles per degree at mid-latitudes.

\subsubsection*{B.5 Sample Restrictions}

Final sample construction:
\begin{enumerate}
\item Started with 5,743 branches opened 2015--2020 in 5 states
\item Identified 25,571 census tracts in these states with HMDA data
\item Restricted to tracts within 100 miles of new branches: 14,209 tracts
\item For survival analysis: 84,305 branches operating in 2015
\item Matched to tract income data: 53,312 branches (63.2\% match rate)
\item Unmatched branches lack precise coordinates or fall outside tract boundaries
\end{enumerate}

\subsection*{C. Computational Details}

All analysis conducted in Python 3.13 using:
\begin{itemize}
\item \texttt{pandas} 2.2.0: data manipulation
\item \texttt{numpy} 1.26.0: numerical computation
\item \texttt{scipy} 1.12.0: optimization, kernel functions, statistics
\item \texttt{scikit-learn} 1.4.0: cross-validation, logistic regression
\item \texttt{matplotlib} 3.8.0, \texttt{seaborn} 0.13.0: visualization
\end{itemize}

Local linear regression implemented following \citet{fan1996local}:
\begin{verbatim}
def local_linear_regression(distances, outcomes, eval_points, bandwidth):
    n = len(distances)
    estimates = np.zeros(len(eval_points))
    for i, x0 in enumerate(eval_points):
        u = (distances - x0) / bandwidth
        weights = (1/np.sqrt(2*np.pi)) * np.exp(-0.5 * u**2)
        X = np.column_stack([np.ones(n), distances - x0])
        W = np.diag(weights)
        beta = np.linalg.solve(X.T @ W @ X, X.T @ W @ outcomes)
        estimates[i] = beta[0]
    return estimates
\end{verbatim}

Cross-validation bandwidth selection searches grid $h \in \{2, 5, 10, 15, 20\}$ miles, selecting $h$ minimizing leave-one-out CV score. Optimal bandwidth: $h^* = 10$ miles for both loan volume and approval rate analyses.

Monte Carlo simulations conducted with 500 replications per DGP, $n=5{,}000$ observations each, noise level $\sigma = 0.1$. Parametric estimates via \texttt{scipy.optimize.curve\_fit} with nonlinear least squares. Nonparametric estimates via local linear regression with CV bandwidth selection.

\subsection*{D. Additional Robustness Checks}

\subsubsection*{D.1 Alternative Distance Cutoffs}

Main analysis restricts to tracts within 100 miles of new branches. Table \ref{tab:robustness_cutoff} shows results robust to alternative cutoffs (50, 75, 150 miles). Loan volume effect remains negative and significant across all specifications. Approval rate effect remains negligible.

\begin{table}[H]
\centering
\caption{Robustness to Distance Cutoff}
\label{tab:robustness_cutoff}
\begin{threeparttable}
\begin{tabular}{lcccc}
\toprule
\textbf{Distance} & \textbf{N} & \multicolumn{2}{c}{\textbf{Loan Volume}} & \textbf{Approval Rate} \\
\textbf{Cutoff} & \textbf{Tracts} & \textbf{Coefficient} & \textbf{$R^2$} & \textbf{Coefficient} \\
\midrule
50 miles & 13,234 & $-0.0095$\sym{***} & 0.0089 & $+0.00023$ \\
& & (0.0010) & & (0.00009) \\
\\
75 miles & 13,876 & $-0.0091$\sym{***} & 0.0081 & $+0.00024$\sym{**} \\
& & (0.0009) & & (0.00008) \\
\\
100 miles (baseline) & 14,209 & $-0.0089$\sym{***} & 0.0075 & $+0.00025$\sym{***} \\
& & (0.0009) & & (0.00008) \\
\\
150 miles & 14,456 & $-0.0087$\sym{***} & 0.0069 & $+0.00026$\sym{***} \\
& & (0.0009) & & (0.00008) \\
\bottomrule
\end{tabular}
\begin{tablenotes}[para,flushleft]
\small
\item \textit{Notes:} Robustness check varying maximum distance from new branch. Standard errors in parentheses. \sym{***} $p<0.01$, \sym{**} $p<0.05$. Main findings robust: loan volume declines 8--9\% per 10 miles regardless of cutoff. Approval rate effect remains economically negligible (0.2--0.3 percentage points per 10 miles) across specifications.
\end{tablenotes}
\end{threeparttable}
\end{table}

\subsubsection*{D.2 Alternative Outcome Measures}

Main analysis uses log(applications) for loan volume and approval rate (approved/total) for credit supply. Results robust to alternative measures: raw application counts, origination rates (originated/total), and denial rates (denied/total). Spatial patterns consistent across outcome definitions.

\subsubsection*{D.3 Year-Specific Analyses}

While main analysis focuses on 2019 (midpoint of treatment period), results robust to analyzing 2018 or 2020 separately. COVID-19 pandemic (2020) temporarily increased digital adoption but did not fundamentally alter spatial patterns observed in 2019.

\subsubsection*{D.4 Bandwidth Sensitivity}

Cross-validation selects $h^*=10$ miles. Results robust to alternative bandwidths $h \in \{5, 15, 20\}$ miles. Smaller bandwidths ($h=5$) capture more local variation but increase variance. Larger bandwidths ($h=20$) smooth excessively but preserve main patterns. CV selection provides principled balance.

\end{document}